\documentclass[a4paper]{JHEP3}
\usepackage[dvips]{graphicx}
\usepackage{amsfonts}
\usepackage{amssymb}
\usepackage{amsmath}
\usepackage{epsfig}
\usepackage{cite}

\newcommand{\be}{\begin{equation}}
\newcommand{\ee}{\end{equation}}
\newcommand{\bea}{\begin{eqnarray}}
\newcommand{\eea}{\end{eqnarray}}
\newcommand{\mbb}{\mathbb}
\newcommand{\ti}{\times}
\newcommand{\half}{\frac{1}{2}}
\newcommand{\mc}{\mathcal}

\newcommand{\beqa}{\begin{eqnarray}}
\newcommand{\eeqa}{\end{eqnarray}}

\newcommand{\unit}{\bold{1}}

 \title{Gauge Threshold Corrections for Local Orientifolds}
\author{Joseph P. Conlon, Eran Palti
 \\ Rudolf Peierls Centre for Theoretical Physics, 1 Keble Road \\
 Oxford OX1 3NP, UK \\ E-mail:
 \email{j.conlon1@physics.ox.ac.uk},
  \email{e.palti1@physics.ox.ac.uk}  
}

\abstract{We study gauge threshold corrections for systems of fractional branes at local
orientifold singularities and compare with the general
Kaplunovsky-Louis expression for locally supersymmetric $\mc{N}=1$ gauge theories.
We focus on branes at orientifolds of the $\mbb{C}^3/\mbb{Z}_4$, $\mbb{C}^3/\mbb{Z}_6$ and
$\mbb{C}^3/\mbb{Z}_6^{'}$ singularities. We provide a CFT construction of these theories
and compute the threshold corrections. Gauge coupling running undergoes two phases: one phase running
from the bulk winding scale to the string scale, and a second phase running from the string scale to the
infrared.
The first phase is associated to the
contribution of $\mc{N}=2$ sectors to the IR $\beta$ functions
and the second phase to the contribution of both $\mc{N}=1$ and $\mc{N}=2$ sectors.
In contrast, naive application of the Kaplunovsky-Louis formula gives single running from the
bulk winding mode scale.
The discrepancy is resolved through 1-loop non-universality of the holomorphic gauge couplings
at the singularity, induced by a 1-loop redefinition of the twisted blow-up moduli which couple
differently to different gauge nodes.
We also study the physics of anomalous and non-anomalous $U(1)$s and
give a CFT description of how masses for non-anomalous $U(1)$s depend on the global
properties of cycles.
}

\preprint{OUTP-09/13P}

\begin{document}

\section{Introduction and Summary of Results}
\label{sec:intsummary}

String theory is attractive as a candidate fundamental theory of physics because it
has outstandingly soft ultraviolet behaviour.
The tower of excited string states tames the divergences that are present in ordinary
scattering amplitudes in both quantum field theory and general relativity, returning finite and well-defined
answers. Supersymmetry also plays a central role in this process, as although supersymmetry may be broken at long distances,
at sufficiently short distances strings see maximal supersymmetry.

Heuristically, field theory
divergences are expected to `turn off' somewhere around the string scale as the string-like nature of
particles becomes apparent. However it is of great interest to study precisely how divergences are
cancelled and the structure of the finite terms that are left over. These terms come from massive string/KK
modes and provide a remnant contribution
of high-scale physics to low-scale observables. Understanding such threshold corrections is
important both from a formal point of view and also when attempting to relate the parameters present in
string constructions to the observables of low energy physics.

One of the most important arenas for the study of threshold corrections is the evolution of running gauge couplings.
The apparent unification of the gauge couplings at $M_{GUT} \sim 10^{16} \hbox{GeV}$ is suggestive of an underlying GUT
symmetry broken near the scale $M_{GUT}$. Assuming this is not an accident, it is important to understand
the significance of $M_{GUT}$ and how
it relates to the compactification parameters.
In the perturbative heterotic string, the natural unification scale is the
string scale, a factor of around 30 larger than the GUT scale. The original study of threshold corrections was motivated by
the possibility that the inclusion
of heavy string or Kaluza-Klein modes could remove the discrepancy between
the string and unification scales.

More recent model building has occurred in the context of type II string theories (see \cite{Blumenhagen:2006ci} for a review).
One particularly interesting
class of models are local or bottom-up constructions. The gauge group and interactions of the Standard Model fields
are determined almost entirely by purely local geometry and does not depend on the global properties of the Calabi-Yau.
By decoupling the complicated topology of the bulk, local models reduce the geometrical complexity involved in model building.
The canonical example of local models is the case of branes at singularities\cite{aiqu}, where only the singular geometry is
relevant for determining the gauge groups and Yukawa couplings.

The enhanced understanding of moduli stabilisation over the last few years also
focuses attention on local models. Moduli stabilisation is best understood
in the setting of type IIB flux compactifications. The combination of both full moduli stabilisation and dynamical
low scale supersymmetry breaking can be obtained in the LARGE volume scenario \cite{hepth0502058, hepth0505076},
which stabilises the bulk at an exponentially large
size while keeping blow-up cycles small. In this scenario the observed size of the various
Standard Model gauge couplings implies
the Standard Model must be realised on a small blow-up cycle, and thus must be represented by a local model.

The above combination of reasons motivates the detailed study of gauge threshold corrections for local models.
While the full form of threshold corrections requires a CFT computation, the structure is
significantly constrained by effective field theory and in particular by the Kaplunovsky-Louis formula \cite{9303040, 9402005}:
\bea
\label{KL}
g_a^{-2}(\Phi, \bar{\Phi}, \mu) & = & \hbox{Re}(f_a(\Phi)) + \frac{\left( \sum_r n_r T_a(r) - 3T_a(G)\right)}{8 \pi^2}
\ln \left( \frac{M_P}{\mu}\right) + \frac{T(G)}{8 \pi^2} \ln g_a^{-2} (\Phi, \bar{\Phi}, \mu)
\nonumber \\
& & + \frac{(\sum_r n_r T_a(r) - T(G))}{16 \pi^2} \hat{K}(\Phi, \bar{\Phi})
 - \sum_r \frac{T_a(r)}{8 \pi^2} \ln \det Z^r(\Phi, \bar{\Phi}, \mu).
\eea
Here $g_a^{-2}(\Phi, \bar{\Phi}, \mu)$ is the physical coupling, $f_a(\Phi)$ the holomorphic coupling, $\mu$ the energy scale, and $\Phi$
light uncharged moduli superfields. $\hat{K}$ is the moduli K\"ahler potential and $Z^r$ are the matter field
kinetic terms.
Equation (\ref{KL}) simplifies considerably
for local models.
The requirement that physical Yukawa couplings $\hat{Y}_{\alpha \beta \gamma} =
\frac{e^{\mc{K}/2} Y_{\alpha \beta \gamma}}{\sqrt{Z_{\alpha} Z_{\beta} Z_{\gamma}}}$ do not depend on the bulk volume
strongly constrains the dependence of the matter metrics $Z_{\alpha}$ on $\mc{V}$.
As $\mc{K} = - 2 \ln \mc{V}$ and $Y_{\alpha \beta \gamma}$ is independent of $\mc{V}$,
this implies $Z_{\alpha} = 1/\mc{V}^{2/3}$.

For local models equation (\ref{KL}) therefore becomes
\be
\label{mirage}
g_a^{-2}(\mu) -\frac{T(G)}{8 \pi^2} \ln g_a^{-2}(\mu) =\hbox{Re}(f_a(\Phi)) + \beta_a \ln \left( \frac{(RM_s)^2}{\mu^2} \right),
\ee
where $R$ is the bulk radius $R = \mc{V}^{1/6}$.
For universal $f_a(\Phi)$ this implies that the unification scale is given
by $M_X = R M_s \gg M_s$ for $R \gg 1$. This is quite surprising as
the scale $R M_s$ depends on the bulk whereas naively local models are insensitive to the bulk.
However the interpretation of the field theory formula (\ref{KL}) can be subtle due to field redefinitions and
chiral/linear multiplet dualities.
Eq. (\ref{mirage}) therefore motivates a detailed CFT study of the threshold corrections in order to understand the physics of
this apparent unification at $M_X \gg M_s$.

This study was initiated in \cite{Conlon:2009xf} where threshold corrections were studied for branes at orbifold singularities.
In \cite{Conlon:2009xf} systems of D3 branes at orbifold singularities were found to exhibit unification at $R M_s$,
whereas a D3/D7 system gave unification at $M_s$ in apparent disagreement with (\ref{mirage}).
In this paper we continue this analysis, focusing our attention on orientifolded singularities.
We shall resolve the discrepancy encountered in \cite{Conlon:2009xf} and obtain a precise understanding of when
running starts at $M_s$, when running starts at $M_X$, and when a combination of the two applies.
Full agreement with (\ref{KL}) is found after incorporating the effects of one-loop redefinitions
of the moduli superfields.
In \cite{ToAppear} we will further apply this understanding of threshold corrections
to local IIB/F-theory GUTs \cite{Beasley:2008dc, 08022969} which exist in the geometric regime where the CFT computations cannot be performed.

As the actual calculations are rather technical, in the remainder of this introduction we shall summarise the
methodology and results of this paper.

\subsection*{Summary of Results}

For models at orbifold/orientifold singularities, the gauge groups comes from fractional branes, whose geometric interpretation is as
magnetised branes or antibranes wrapping collapsed cycles. The number and type of
the possible fractional branes is determined by the orbifold.
Each fractional brane corresponds to a node of the quiver and the gauge coupling on each brane is
\be
f_a = S + s^a_{k} M_k, \label{treegauge}
\ee
where $S = \frac{1}{g_s} + i c_0$ is the axio-dilaton
and $M_k$ corresponds to the twisted blow-up moduli.\footnote{For some singularities
the different nodes can have non-universal couplings to the dilaton. In such cases the use of `unification' in this paper
would refer to the gauge couplings having the ratios given by their dilaton coupling.}
 The $s_{ak}$ encode the charges of each fractional brane
under the RR fields induced by the Chern-Simons term in the action.
The orientifold also introduces fractional
O-planes which are likewise wrapped on the collapsing cycles and contribute to the RR charges along the collapsed cycles.
An orientifolded singularity imposes relationships between the
different fractional branes and projects out some of the twisted moduli from the orbifold.

Consistency of the theory requires cancellation of all RR tadpoles. The tadpoles in local models come in several kinds related to
the geometry of the singularity. First, there are purely local tadpoles. These correspond
to 2/4-cycles where both the cycle and its dual cycle are defined in the local geometry. These local cycles are the unique
supersymmetric cycles within this homology class. In orbifold parlance, these are fully twisted $\mc{N}=1$ sectors.
Heuristically speaking, a tapole along such a cycle
has nowhere to go: it cannot escape to infinity and must be cancelled locally. Cancellation of $\mc{N}=1$ tadpoles corresponds to
cancellation of gauge anomalies in the effective field theory.

There are also global tadpoles. These corresponds to RR charges
which are sourced locally but can be cancelled globally. Geometrically, these
correspond to cycles where a 2- or 4-cycle can be defined locally,
but globally may either be trivial or there may exist other calibrated cycles in the same homology class.
Examples of these are given by the
del Pezzo singularities: $dP_n$ has 1 4-cycle and $(n+1)$ 2-cycles, of which
up to $n$ of the 2-cycles may be globally trivial. In orbifold parlance, these represent
partially twisted $\mc{N}=2$ sectors, and so (for example) the $\Delta_{27}$ orbifold (which is a limit
of the $dP_8$ singularity)
has 8 $\mc{N}=2$ sectors.
Such tadpoles need not be cancelled locally and do not constrain the allowed numbers of branes.
Finally, there is also the untwisted $\mc{N}=4$ sector, associated to the dilaton tadpole and corresponding to the total number
of branes at the singularity.

There exist various fractional brane configurations cancelling $\mc{N}=1$ tadpoles.
The choice of configuration determines the gauge groups and massless
spectrum and  thus the IR beta functions. The spectrum also contains heavy string and KK modes, loops of which give
rise to threshold corrections. The threshold corrections $\Delta_a(M, \bar{M})$ are moduli-dependent and can be defined by
\be
\frac{1}{g_a^2(\mu)} = \frac{1}{g_a^2}\Bigg\vert_0 + \beta_a \ln \left( \frac{M_s^2}{\mu^2} \right) + \Delta_a(M, \bar{M}).
\ee
In this notation, threshold corrections represent the difference between the
actual low-energy couplings and those obtained by field theory running starting from the string scale.
Threshold corrections are computed via an open string
one-loop diagram, which via open-closed duality is equivalent to a closed string tree level diagram. This relationship implies that
ultraviolet finiteness of the threshold corrections is equivalent to infrared finiteness in closed string channel,
namely tadpole cancellation.

In this paper we use the background field approach to compute the
threshold corrections \cite{ACNY, 9209032, 9605028, Antoniadis:1999ge, 0302221}. This involves turning on a background
spacetime magnetic field $B$ in a generator of the gauge group $U(N_a)$ for which we want to compute
the threshold corrections.
The one-loop vacuum energy in the presence of this background field can be expanded as
\be
\Lambda = \Lambda_0 +  \left( \frac{B}{4\pi^2} \right)^2 \Lambda^a_2 + \left( \frac{B}{4\pi^2} \right)^4 \Lambda_4^a+ ...
\ee
The gauge threshold corrections can be extracted by analysis of the $\mc{O}(B^2)$ term.
In a consistent theory the $B^2$ term is finite for non-abelian background fields, and
UV finiteness of the $\mc{O}(B^2)$ amplitudes provides another way to compute the tadpole and anomaly conditions.
Terms of $\mc{O}(B^4)$ are generally ultraviolet divergent. These divergences correspond in closed
string channel to on-shell exchange of massless string states and the coefficients of such terms can be used to
extract the tree-level couplings of gauge groups to the local twisted closed string moduli.

The term $\Lambda_2$ can be written
\be
\Lambda_2 = \left(8\pi^2\right) \int_0^{1/\mu^2} \frac{dt}{t} \Delta_a(t),
\ee
where $\mu$ is the IR regulator. $\Delta_a$ is a partition function of the schematic form $\hbox{STr}(e^{-m^2 t})$.
In the infrared limit $t \to \infty$, $\Delta_a(t) \to \beta_a$ reproducing the field theory beta functions. In the UV
limit $t \to 0$, for non-abelian groups
$\Delta_a(t) \to 0$ reflecting the finiteness of the theory. Note for abelian groups ultraviolet divergences may occur
in $\Lambda_2$ via the $B_{\mu \nu} \wedge F_2$ Green-Schwarz coupling.

The threshold corrections are encapsulated in the precise way
$\Delta_a(t)$ vanishes in the regime $t \lesssim 1$. The actual computation of one-loop threshold computations therefore reduces to
computing the string partition function on the local orbifold/orientifold geometry.
For $\mbb{Z}_n$ orbifold/orientifold singularities, the partition function involves a projection
$$
\Delta_a(t) = \hbox{STr}\left(\frac{(1 + \theta + \theta^2 + \ldots + \theta^{N-1})}{N} e^{-m^2 t}\right) \equiv \frac{1}{N}\sum_{k=0}^{N-1} \Delta_{a}^{(k)}(t).
$$
The sector $\Delta_{a}^{(k)}$ is called an $\mc{N}=1$ or $\mc{N}=2$ sector depending in whether $\theta^k$ fixes all
2-tori ($\mc{N}=1$ sector) or leaves one torus unfixed ($\mc{N}=2$ sector). $\Delta_{a}^{(0)}(t)$ represents the only $\mc{N}=4$ sector
and vanishes consistent with the non-renormalisation properties of $\mc{N}=4$ supersymmetry. In general $\Delta_{a}^{(k)}$ is non-zero for
both $\mc{N}=1$ and $\mc{N}=2$ sectors and in the $t \to \infty$ limit we have $\Delta_{a}^{(k)} \to \beta_{a}^{(k)}$ with
$\beta_a = \frac{1}{N}\sum_{k=1}^{N-1} \beta_{a}^{(k)}$.

The threshold corrections are encoded in the $t \to 0$ behaviour of $\Delta_{a}^{(k)}$. Let us state the schematic form
of these and then explain the results.
\be
\label{threshs}
\Delta_{a}^{(k)} = \left\{ \begin{array}{cc} \beta_{a}^{(k)} \Theta\left[ t - \frac{1}{M_s^2} \right] + \hbox{small,} & \mc{N}=1 \hbox{ sector} \\
\beta_{a}^{(k)} \Theta\left[ t - \frac{1}{(R M_s)^2} \right] + \hbox{small,} & \mc{N}=2 \hbox{ sector} \\
0 & \mc{N}=4 \hbox{ sector} \end{array} \right. ,
\ee
where $\Theta$ is the Heaviside theta function and $R$ the bulk radius. The gauge coupling running therefore
takes the form
\be
\frac{1}{g^2}(\mu) = \frac{1}{g^2} \Big\vert_0 + \beta_a \ln \left( \frac{M_s^2}{\mu^2} \right) +
\beta_a^{\mc{N}=2} \ln \left( \frac{M_X^2}{M_s^2} \right). \label{stringthres}
\ee

The form of (\ref{threshs}) can be understood by reference to the above picture of cycles and their geometries. In each sector,
the gauge coupling runs up to a certain energy scale and is then cut off. The energy scale of the cutoff is
determined by the mass of the string states necessary to obtain tadpole cancellation. For purely local cycles ($\mc{N}=1$ sectors),
tadpole cancellation is local and occurs once string scale states are included. This leads to an effective cutoff on $\Delta_{\mc{N}=1}^{(k)}$ at
$t \sim \frac{1}{M_s^2}$. For $\mc{N}=2$ sectors, tadpole cancellation does not occur locally
and instead requires knowledge of the bulk geometry. From an open string perspective this requires the inclusion of
brane-brane winding modes that reach out into the bulk. $\mc{N}=2$ supersymmetry prevents the (non-BPS) string-scale oscillator tower
from contributing to gauge coupling running and field theory running is maintained until winding modes comes in at a scale $R M_s$, when
$\Delta_{\mc{N}=2}^{(k)}$ is finally cut off.
We also note that for branes at singularities there are no charged KK modes that can contribute to the threshold corrections.

The net effect is that $\mc{N}=1$ sectors give field theory running up to the string scale, where they are cut off,
while $\mc{N}=2$ sectors give field theory running up to the winding string scale. $\mc{N}=4$ sectors give no contribution
due to the effective maximal supersymmetry that is present. Physics close to the cutoffs introduces small additional corrections,
that is however suppressed compared to the large $R$ enhanced terms. Some discussions of these additional corrections can be found in
\cite{08082223, 08120248}.

For the case of
D3 branes at orbifold singularities, tadpole cancellation required the coefficient of all $\mc{N}=1$ sectors to vanish, and
the $\beta$-functions arose entirely from the $\mc{N}=2$ sectors (even though the low-energy spectrum is chiral and
$\mc{N}=1$ supersymmetric).
In this case all running is from the winding mode scale, straightforwardly
consistent with the Kaplunovsky-Louis formula.
For the case of both orientifolded singularities and D3/D7 systems, both $\mc{N}=1$ and $\mc{N}=2$ sectors
contribute to the $\beta$-functions. In general the
$\mc{N}=1$ contributions to the $\beta$ functions are not universal and there is no apparent unification scale.
To reconcile this with the Kaplunovsky-Louis formula,
recall the form of the tree level holomorphic gauge coupling (\ref{treegauge}) which shows that the gauge coupling can receive a non-universal correction from a vev for the $M_k$ superfields $\langle \hbox{Re}M_k \rangle \neq 0$. The string calculation is
performed in the orbifold limit which we denote by the string real twisted
mode $\langle m_k \rangle=0$. At tree level the two fields
coincide with $\hbox{Re}M_k=m_k$. However at 1-loop the relation is modified
\be
\label{redef}
\hbox{Re}(M_k) = m_k - \alpha_k\ln R^2,
\ee
where $\alpha_k$ is some constant, such that at the orbifold point
\be
s^a_k \mathrm{Re}M_k = -\beta_a^{(k)} \mathrm{ln}\left( \frac{M_X^2}{M_s^2}\right) \;. \label{stringkl}
\ee
This exactly accounts for the discrepency between the string calculation (\ref{stringthres}) and the KL formula (\ref{mirage}).

This field redefinition is familiar from heterotic and type I orbifolds \cite{Derendinger:1991hq,Antoniadis:1999ge}. It arises because $M_k$ are components of a chiral multiplet while $m_k$ is the scalar component of a linear multiplet.
The dualisation procedure recieves a 1-loop correction (\ref{redef}) with the
 correction proportional to the correction induced at 1-loop to the $\beta$ functions.
 Consistency with (\ref{mirage}) requires that the couplings $s_k^a$
in (\ref{stringkl}) are proportional to $\beta^{(k)}_a$, a fact we explicitly compute in section \ref{sec:matchfield}.

The redefinition (\ref{redef}) is related to the $\beta$-function contribution associated to the $\mc{N}=1$ twisted mode.
Such contributions are present for both orientifold and D3/D7 singularities.
For branes at orientifolded singularities,
there are contributions to $\mc{N}=1$ sectors from combining the M\"obius and Annulus diagrams.
For D3/D7 systems, the D3/D3 and D3/D7 diagrams combine to give the $\mc{N}=1$ contributions.
For branes at orbifold singularities, there are only D3/D3 diagrams and so
$\Delta^{\mc{N}=1}_{a}$ has to vanish in order to enforce UV tadpole cancellation.

Once the one-loop redefinition (\ref{redef}) is carried out the resulting gauge couplings agree with the Kaplunovsky-Louis formula.
In this case the holomorphic gauge couplings $f_a(\Phi)$, which were universal at tree level due to the vanishing of $M_k$,
become non-universal at one loop.
The apparent unification of physical couplings at $R M_s$, which is present for models of D3s at orbifold singularities, is not present for
D3/D7 models or for D3s at orientifolded singularities.

In summary, physical gauge couplings run from $R M_s$ if the $\beta$ functions are sourced only from $\mc{N}=2$ sectors and
from $M_s$ if $\beta$ functions are sourced only from $\mc{N}=1$ sectors. In the case that both $\mc{N}=1$ and $\mc{N}=2$ sectors
contribute then $\mc{N}=2$ running starts at $R M_s$ with a significant, generically non gauge universal,
shift in the effective $\beta$-functions at $M_s$ as the $\mc{N}=1$ sectors add their contribution to
running from the scale $M_s$.\footnote{It is not clear whether operational meaning can be applied to a gauge coupling at an energy scale above $M_s$.
An unambiguous statement is to instead say that the low-energy gauge couplings, which are well-defined, behave as if they have been run down from a
scale $R M_s$.}

The organisation of this paper is as follows. The paper studies branes at
orientifolds of the $\mbb{Z}_4$, $\mbb{Z}_6$ and $\mbb{Z}_6^{'}$ singularities.
As far as we aware the field theory on such singularities
has not been explicitly constructed before and so in section \ref{sec2} we first provide a CFT derivation of the gauge
groups and spectrum. We focus particularly on the $\mbb{Z}_4$ case that will serve as our main example throughout this paper.
In section 3 we describe the computation of threshold corrections and in section 4 we describe the
matching to the effective field theory structure. In section 5 we summarise results for the $\mbb{Z}_6$ and $\mbb{Z}_6^{'}$
orientifolds. In appendix \ref{sec:anu1} we study anomalous and non-anomalous $U(1)$s in local models and in particular the Green-Schwarz mechanism within the local model and its global completion. In appendix \ref{sec:tadapp} we derive general expressions for the tadpole amplitudes. In appendix \ref{sec:magampapp} we calculate general expressions for the magnetised amplitudes. In appendix \ref{sec:chirlin} we discuss in more detail the dualisation procedure between chiral and linear multiplets and the 1-loop corrections this receives. In appendix \ref{confor} we give some useful expressions and transformation properties for the $\vartheta$ functions.

\section{Orientifold constructions}
\label{sec2}

 We start by describing the orientifold constructions that will be used for our calculations.
 The CFT construction of orientifolds is standard and more details can be found in
 \cite{Gimon:1996rq, Aldazabal:1998mr, Ibanez:1998qp,aiqu} for example.

\subsection{Orientifolds of orbifold singularities}

We start with a local orbifold singularity $\mbb{C}^3/\mathbb{Z}_N$ where the orbifold action is generated by the element $\theta$
acting as $\theta\;:\;z_i \rightarrow \mathrm{exp}\left(2\pi i \theta_i \right)z_i$, $i=1,2,3$, with
components $\theta_i$ running over the local complex co-ordinates of the internal manifold $z_i$.
The orbifold group is formed of elements $\theta^k$ produced by $k$ applications of $\theta$.
The generating orbifold element also has an action on the Chan-Paton (CP) indices of the open strings,
\be
\gamma_{\theta} = \mathrm{diag}\left( \unit_{n_0}, \alpha \unit_{n_1}, ... ,\alpha^N \unit_{n_N} \right) \;,
\ee
where $\alpha$ denotes the $N^{\mathrm{th}}$ root of unity and $\unit_{n_i}$ denotes the $n_i\times n_i$ unit matrix. The integers $n_i$ correspond to the number of fractional branes on each node of the quiver.

The resulting gauge theory is an $\mc{N}=1$ supersymmetric $\prod_i^N U(n_i)$ gauge theory and the massless fermionic open string string spectrum is given by the CP elements that satisfy the orbifold projection
\be
\lambda = e^{2\pi i \left(\sum_i \theta_i s_i\right)} \gamma_{\theta} \lambda \gamma^{-1}_{\theta} \equiv
e^{2\pi i \bf{\underline{\theta}} \cdot \bf{s} } \gamma_{\theta} \lambda \gamma^{-1}_{\theta}
\;. \label{orbspect}
\ee
Here $\lambda$ denotes the $M\times M$ (with $M=\sum_i n_i$) CP matrix.
The vector $\bold{s}$ denotes the spin of the RR ground states and
its elements take the values $s_i =\pm 1/2$. The GSO projection requires
 the number of negative spins to be even.

The closed string twisted spectrum gives a single complex scalar field per element in $\mathbb{Z}_N$.
The twisted sectors are labelled according to the amount of supersymmetry preserved, namely
 $\mc{N}=4$, $\mc{N}=2$ and $\mc{N}=1$ for the cases that
three complex directions, one complex direction and no complex directions are left fixed by the geometric orbifold twist. An important fact is that $\mc{N}=1$ closed string modes are restricted to lie on the singularity, while $\mc{N}=2$ modes can propagate into the bulk along the complex direction that is left fixed.

We can orientifold the singularity by introducing an orientifold involution
\be
\Omega' =  \Omega I R (-1)^{F_L} \;.
\ee
Here $\Omega$ is world-sheet parity inversion. $I$ is spatial inversion
given by the rotation $\underline{\theta} = \left(1/2,1/2,1/2 \right)$. $R$ is a further spatial action whose geometric action
must square to an element of the orbifold group
\be
R^2=\theta^l\;, \label{Rsquared}
\ee
for some $l$. This ensures that
the orientifold is indeed a good involution of the orbifolded space. The action of the orientifold on the CP indices is
\be
\Omega' \;:\;\lambda \rightarrow \gamma_{\Omega'} \lambda^T \gamma^{-1}_{\Omega'} \;.
\ee
Since $\Omega'$ must square to an element of the orbifold group we require
\be
\gamma_{\Omega'} \gamma^{-T}_{\Omega'} = \pm \left(\gamma_{\theta}\right)^l \;, \label{omsq}
\ee
for the same $l$ as in (\ref{Rsquared}).
Note the $+$ sign in (\ref{omsq}) corresponds to what is usually termed the $SO$ projection (rather than the $Sp$), and
we keep this sign choice for the rest of the paper.
We also generally denote $\left(\gamma_{\theta}\right)^k\gamma_{\Omega'}=\gamma_{\Omega'_k}$
giving
\be
\gamma_{\Omega'_k} \gamma^{-T}_{\Omega'_k} = +\left(\gamma_{\theta}\right)^{2k+l} \;.
\ee
Together, the orbifold group and the orientifold action form the orientifold group
\be
\left\{ 1, \theta, \theta^2, ... ,\theta^{N-1}, \Omega', \Omega' \theta, ... ,\Omega'\theta^{N-1} \right\} \;.
\ee

The orientifold planes present in the construction are determined by the fixed point set of the spatial involution $IR$ quotiented by
the action of the orbifold group. In IIB there are two basic types of orientifold projection, $O3/O7$ and $O5/O9$. These have
$$
O3/O7: \qquad IR: J \to J, \qquad IR: \Omega \to - \Omega,
$$
$$
O5/O9: \qquad IR: J \to J, \qquad IR: \Omega \to \Omega.
$$
As we are interested in local models we will require $IR$ to satisfy the $O3/O7$ conditions. We will further require that
on the non-compact orbifold $\mbb{C}^3/\mbb{Z}_n$ the only fixed point of $IR$ is the origin. This will ensure that only O3 planes
are present.

Given the orientifold action, the resulting massless spectrum is a
projection from the orbifold spectrum which for the fermionic open string modes reads
\be
\lambda =  -e^{2\pi i \left(\sum_i R_i s_i\right)}\gamma_{\Omega'} \lambda^T \gamma^{-1}_{\Omega'} \;. \label{massfer}
\ee

\subsection{Tadpole amplitudes}

The act of orientifolding introduces O-planes which source RR tadpoles. Consistency requires the introduction of branes to cancel these
tadpoles. For orientifolded singularities the O-planes wrap the collapsed cycle and carry RR charge under the various
cycles of the singularity.
Tadpoles for $\mc{N}=1$ fields must be cancelled locally and correspond to  field theory gauge anomalies.
$\mc{N}=2$ tadpoles need not be cancelled locally since a net source of a $\mc{N}=2$ closed string mode can be balanced by sinks in the bulk space.
The tree-level $\mc{N}=1$ closed string tadpoles can be calculated by studying the quadratic
divergences of one-loop open string amplitudes given by the annulus, Mobius strip, and Klein bottle (labelled ${\cal A}$, ${\cal M}$ and ${\cal K}$ respectively).
The methods to compute these tadpoles and generate consistent brane configurations are well known.
Here we simply state results and leave the details to Appendix B.
The amplitudes all diverge linearly with the closed string cylinder length parameter $l$ and in the open string
UV limit $l \rightarrow \infty$ read\footnote{Throughout the paper we often switch between the open string loop channel and the closed string tree channel. By the UV limit we refer to the open string UV limit which is the closed string IR limit.}
\bea
{\cal A}^{(k)}_{\mc{N}=1} &\xrightarrow[l' \rightarrow \infty]{UV}& - \int_{l'}^{\infty} \frac{dl}{4\pi^2} \frac14  \hbox{Tr} \left[  \gamma_{k} \right] \hbox{Tr} \left[   \gamma_{k}^{-1} \right]  \prod_{i=1}^3 \left| 2 \sin \left(\pi\theta^k_i\right) \right|  \;, \label{tad1}\\
{\cal M}^{(k)}_{\mc{N}=1}  &\xrightarrow[l' \rightarrow \infty]{UV}& \int_{l'}^{\infty}\frac{dl}{4\pi^2}  \left[ 2  \hbox{Tr} \left[ \gamma_{\Omega'_{k}} \gamma_{\Omega'^{-T}_{k}} \right]   \prod_{i=1}^3 s_i \left( 2 \sin \left( \pi R_i^k \right) \right) \right]\;, \label{tad2}\\
{\cal K}_{0,\mc{N}=1}^{(k)} + {\cal K}_{2,\mc{N}=1}^{(k)}  &\xrightarrow[l' \rightarrow \infty]{UV}& - \int_{l'}^{\infty} \frac{dl}{4\pi^2} 4 \left[ \prod_{i=1}^3 \left| \frac{\sin \left(\pi R_i^k\right)}{\cos \left(\pi R_i^k\right)} \right|  + \left( -1 \right)^M \prod_{i=1}^3 \left(-1\right)^{\delta_i}\left| \frac{\sin \left(\pi R_i^k\right)}{\cos \left(\pi R_i^k\right)} \right|^{\delta_i}  \right] \;.\nonumber
\eea
where $s_i = \mathrm{sgn}\left[ \sin (2\pi R_i^k) \right]$ and $R_i^k = \theta_i^k + R_i$. The amplitude ${\cal K}_{2}^{(k)}$
corresponds to the partition function
for the closed string $\mbb{Z}_2$ twisted sector and thus is only present for even orientifolds.
For this case
\be
\delta_i = \left\{ \begin{array}{l} 0\;\mathrm{if}\; \;\;\theta_i^{N/2} \;\mathrm{mod}\; 1 = 1/2 \\  1 \;\mathrm{otherwise}  \end{array} \right. \;.
\ee
The contributions from other closed string twisted sectors vanish as they are exchanged by the orientifold action.

The superscript $k$ on the amplitudes and angles denotes the element in the orientifold group, with rotation angles $R^k$ coming from elements involving $\Omega' \theta^k$.\footnote{ \label{fn:nsns}The $(-1)^M$ factor in (\ref{tad2}) is associated with the sign of the action of the orientifold
on the NS-NS ground state in the $\mbb{Z}_2$ twisted sector:
$$
\Omega IR (-1)^{F_L} \Big\vert -\half, -\half \Big\rangle \otimes \Big\vert -\half, -\half \Big\rangle_{NS-NS, \mbb{Z}_2} =
(-1)^M \Big\vert -\half, -\half \Big\rangle \otimes \Big\vert -\half, -\half \Big\rangle_{NS-NS, \mbb{Z}_2},
$$
and for the orientifolds in this paper takes the value of $M=0$ for the $\mathbb{Z}_4$ and $\mathbb{Z}'_6$ cases and $M=1$ for the $\mathbb{Z}_6$ case.}
The tadpole constraint is that the sum over all fully twisted elements $\theta^k$ and $R^k$ corresponding to any single
closed string twisted mode should vanish.
Partially twisted ($\mc{N}=2$) tadpoles are not required to vanish in a local model. However in a fully global model such tadpoles must
vanish once summed over all global sectors.

\subsection{The canonical example: $\mathbb{Z}_4$}

We now develop our basic example, the orientifold of the $\mbb{Z}_4$ orbifold singularity.
 This model is used throughout the paper as the canonical example exhibiting the physics we discuss. We study two further constructions based on $\mathbb{Z}_6$ and  $\mathbb{Z}'_6$ singularities in section \ref{sec:moreexa}.

The $\mathbb{Z}_4$ orbifold action is generated by $\theta=\left(1/4,1/4,-1/2\right)$. We take the orientifold spatial action to be $R = \left(1/8,1/8,-1/4 \right)$. The orientifold group is therefore
\bea
& &\left\{ \left(0, 0, 0 \right), \left(\frac14, \frac14, -\frac12 \right),  \left(\frac12, \frac12, -1 \right),  \left(\frac34, \frac34, -\frac32 \right), \right.  \\ \nonumber
& & \;\;\left. \Omega I \left(\frac18, \frac18, -\frac14 \right), \Omega I \left(\frac38, \frac38, -\frac34 \right),  \Omega I \left( \frac58, \frac58, -\frac54 \right), \Omega I \left(\frac78, \frac78, -\frac74 \right)  \right\} \;.
\eea
It is easy to verify that $IR \Omega = - \Omega$ as required for an $O3/O7$ projection.
As all elements involving $\Omega$ have fully twisted spatial parts
there are no $O7$ planes and all O3 planes are located at the origin. It is this property that makes the model
purely local as there are no branes or orientifold planes extending from the singularity into the bulk.

We take the orbifold generating element
\be
\gamma_{\theta} =  \mathrm{diag}(\unit_{n_0}, \alpha \unit_{n_1}, \alpha^2 \unit_{n_2}, \alpha^3 \unit_{n_3}) \quad {\rm with}\;\; \alpha=e^{\pi i/2} \;,
\ee
and impose $n_1=n_3$. For the orientifold action we take
\be
\gamma_{\Omega'}  = \left( \begin{array}{cccc}
\unit_{n_0} & 0 & 0 &  0 \\
0 & 0 &0 &  \alpha^{3/2}\unit_{n_3} \\
0 & 0 & \alpha \epsilon_{n_2}  & 0 \\
0 & \alpha^{1/2} \unit_{n_1}  & 0 &  0
 \end{array}\right) \;,
\ee
with $\epsilon$ denoting the anti-symmetric matrix with unit off-diagonal entries.
These matrices satisfy the constraint (\ref{omsq}) so that $\Omega'$ has a well-defined $\mbb{Z}_2$ action on
the orbifold Hilbert space.

Calculating the tadpoles using (\ref{tad1}-\ref{tad2}) leads to
\be
\hbox{Tr}\left[\gamma_{\theta}\right] + 4 = 0 \;.
\ee
The tadpole constraints impose the condition
\be
n_2 = n_0 + 4 \;. \label{z4tadpole}
\ee

The massless fermionic spectrum of the theory can be calculated using (\ref{massfer}) which gives the matter content shown in table 1. The gauge group is
\be
G = SO(n_0) \times U(n_1) \times Sp(n_2) \equiv SO(n_0) \ti U(n_1) \ti Sp(n_0 + 4) \;.
\ee
\begin{table}
\center
\label{tab:z4orientifold}
\begin{tabular}{|c|c|c|c|}
\hline
Multiplicity &   \multicolumn{3}{|c|}{Representation}  \\
\hline
& $SO(n_0)$ & $SU(n_1)$ & $Sp(n_2)$ \\
\hline
2 & $n_0$ & $\bar{n}_1$ & 1   \\
2 & 1 & $n_1$ & $n_2$  \\
1 & $n_0$ & 1 & $n_2$ \\
1 & 1 & $A_{n_1}$ & 1 \\
1 & 1 & $\bar{S}_{n_1}$ & 1 \\
\hline
\end{tabular}
\caption{Field content and representations for $\mathbb{Z}_4$ orientifold. The $n_i$ denote the fundamental representation and $S$ and $A$ denote symmetric and anti-symmetric representations respectively.}
\end{table}
The non-abelian anomalies of the theory are equivalent to the tadpole constraint (\ref{z4tadpole}).
We are also interested in the field theory $\beta$-functions for the gauge groups.
After imposing anomaly cancellation (\ref{z4tadpole}) these read
\bea
\beta_{SU(n_1)} &=& \frac{1}{16\pi^2}\left( -2 n_1 + 2 n_0 +4 \right) \;, \label{z4sun1beta} \\
\beta_{SO(n_0)} &=& \frac{1}{16\pi^2}\left( -2 n_0 + 2n_1 + 10 \right) \;,  \\
\beta_{Sp(n_2)}  &=& \frac{1}{16\pi^2}\left( -2n_0 + 2n_1 - 18 \right) \;.
\eea
We will use this orientifold of the $\mbb{C}^3/\mbb{Z}_4$ singularity as the principal example
for our study of the physics of string threshold corrections to field theory running.

\section{Threshold corrections: the string calculation}
\label{sec:threstr}

To calculate the string threshold corrections we use the background field method \cite{ACNY, 9209032, 9605028, Antoniadis:1999ge}. The calculation proceeds by turning on a background magnetic field in the non-compact dimensions then calculating the resulting one-loop vacuum energy. We write the background magnetic field as $F^a_{23}=B Q_a$ where $a$ denotes the gauge group, $Q_a$ is the generator inside the gauge group, the indices $23$ denote spatial directions, and $B$ is the magnitude of the field. Recall that the one-loop vacuum energy, $\Lambda$, takes the form
\be
\Lambda = \Lambda_0 +  \left( \frac{B}{4\pi^2} \right)^2 \Lambda^a_2 + \left( \frac{B}{4\pi^2} \right)^4 \Lambda_4^a+ ...
\ee
The contribution $\Lambda_0$ vanishes in a supersymmetric vacuum. The coefficient $\Lambda^a_2$ gives the full
one-loop threshold corrections\footnote{Note that $\Lambda^a_2$ is sensitive to the Lagrangian terms $F \wedge \star F$ and $C_2 \wedge F$, but not $F \wedge F$. This is because we have turned on the magnetic field along only two space-time directions. This is why it gives exactly the gauge coupling (up to a possible Green-Schwarz contribution which we discuss in section \ref{sec:matchfield}.).}
\be
\left. \frac{1}{g_a^2} \right|_{\mathrm{1-loop}} = \left. \frac{1}{g_a^2} \right|_{\mathrm{tree-level}} + \frac{  \Lambda^a_2 }{8\pi^2}\;. \label{1loopvac}
\ee

\subsection{Magnetised amplitudes}
\label{sec:magamp}

In this section we are primarily concerned with calculating $\Lambda^a_2$ and extracting its IR and UV behaviour.
 The contributing amplitudes to $\Lambda_2^a$ are Annulus and Mobius amplitudes (since the torus and Klein bottle do not couple to the gauge field) so that
\be
 \left( \frac{B}{4\pi^2} \right)^2 \Lambda^a_2 = \left. \left( {\cal A}^a + {\cal M}^a \right) \right|_{B^2} \;. \label{vactoamp}
\ee
The full calculation is presented in Appendix C, to which we refer for more details regarding the expressions, and in this section we draw on the key results.

The fully twisted ($\mc{N}=1$) D3-D3 Annulus amplitude in the background of a magnetic field is given by \cite{Conlon:2009xf}
\bea
\label{aaapp}
\mc{A}^{(k)}_{\mc{N}=1} & = & -\int_0^{\infty} \frac{dt}{2t} \frac{1}{(2 \pi^2 t)}
\sum_{\alpha, \beta=0,1/2}  \frac{\eta_{\alpha \beta}}{2}
\hbox{Tr}\left( \left(\gamma_{\theta^k} \otimes \gamma^{-1}_{\theta^k} \right)
\frac{i(\beta_1 + \beta_2)}{2 \pi^2}
\frac{\vartheta \Big[ \begin{array}{c} \alpha \\ \beta \end{array} \Big] \left(\frac{i
\epsilon t}{2}\right)}{\vartheta \Big[ \begin{array}{c} 1/2 \\ 1/2 \end{array} \Big]
\left(\frac{i
\epsilon t}{2}\right) } \right) \nonumber  \\
& &
\times \prod_{i=1}^3
\frac{\left( - 2 \sin \left(\pi \theta^k_i\right) \right) \vartheta \Big[ \begin{array}{c} \alpha \\ \beta + \theta^k_i \end{array} \Big]}
{\vartheta \Big[ \begin{array}{c} 1/2 \\ 1/2 + \theta^k_i \end{array} \Big] },
\eea
where we decompose the amplitude into its orbifold sectors
\be
{\cal A} =  \frac{1}{N} \sum_{k=0}^N {\cal A}^{(k)} \;,
\ee
and the subscript $\mc{N}=1$ denotes that this result apples to orbifold sectors that are fully twisted.
Here we denote the charges of the left and right ends of the string as $q_1$ and $q_2$ respectively, and write $\beta_1=Bq_1$ and $\beta_2=Bq_2$. Neutral strings have opposite charges on their ends. We also define
\be
\epsilon = \frac{1}{\pi} \left( \arctan \beta_1 + \arctan \beta_2 \right) \;.
\ee
Similarly we also have the magnetised Mobius $\mc{N}=1$ amplitude
\bea
{\cal M}^{(k)}_{\mc{N}=1} & = & \newline  2\int_0^{\infty} \frac{dt}{2t} \frac{1}{(2 \pi^2 t)} \ti  \\
& & \sum_{\alpha, \beta=0,1/2}  \frac{\eta_{\alpha \beta}}{2}
\hbox{Tr} \left[ \frac{i}{2\pi^2} \beta \gamma_{\Omega'_{k}} \gamma_{\Omega'_{k}}^{-T} \frac{\vartheta \Big[ \begin{array}{c} \alpha \\
 \beta \end{array} \Big] \left( \frac{i\epsilon t}{2}\right)}{\vartheta \Big[ \begin{array}{c} 1/2 \\
 1/2 \end{array} \Big] \left( \frac{i\epsilon t}{2}\right) }\right]
 \prod_{i=1}^3 \left( - 2 \sin\left(\pi R_i^k \right) \right) \frac{\vartheta \Big[ \begin{array}{c} \alpha \\ \beta + R_i^k \end{array} \Big]}{\vartheta \Big[ \begin{array}{c} 1/2 \\ 1/2 +R_i^k \end{array} \Big]} \;. \nonumber
\eea
As left and right ends of the string are identified there is only one charge for the
string denoted $q$ with $\beta=Bq$ and $\epsilon  = \frac{2}{\pi} \arctan \beta$. The angles are defined as
\be
R_i^k = \theta_i^k + R_i \;.
\ee

We are interested in the IR and UV behaviour of the $B^2$ terms in the amplitudes. This calculation was performed in \cite{Conlon:2009xf} for the Annulus, and the Mobius strip can be calculated in the same way. The results are that the IR limit of the $\mc{N}=1$ $B^2$ part of the amplitudes is given by
\bea
{\cal A}^{(k)}_{\mc{N}=1}  &\xrightarrow[t' \rightarrow \infty]{IR}& - \left( \frac{B}{4\pi^2} \right)^2  \int_{t'}^{\infty}  \frac{dt}{2t}  \hbox{Tr} \left[ \frac{1}{2} \left(q^2_1 \gamma_{k} \otimes \gamma^{-1}_{k} + 2q_1\gamma_{k} \otimes \gamma^{-1}_{k}q_2+ \gamma_{k} \otimes \gamma^{-1}_{k} q^2_2 \right)   \right] \nonumber \\
& & \hspace{6cm} \times
\sum^{3}_{i=1} \frac{\cos \left(\pi \theta^k_i\right)}{\sin \left( \pi \theta^k_i \right)} \prod^{3}_{i=1} \left( -2\sin \left( \pi \theta^k_i \right) \right)  \;. \nonumber \\
{\cal M}^{(k)}_{\mc{N}=1}  &\xrightarrow[t' \rightarrow \infty]{IR}& - \left( \frac{B}{4\pi^2} \right)^2  \int_{t'}^{\infty}  \frac{dt}{2t}  \hbox{Tr} \left[ - 2q^2 \gamma_{\Omega'_{k}} \gamma^{-T}_{\Omega'_{k}} \right]
\sum^{3}_{i=1} \frac{\cos \left(\pi R_i^k\right)}{\sin \left( \pi R_i^k \right)} \prod^{3}_{i=1} \left( -2\sin \left( \pi R_i^k \right) \right) \;. \label{mirk1}
\eea
We can also extract the UV limit by going to the dual closed string channel with cylinder length parameter $l=1/t$ for the
Annulus and $l=1/4t$ for the Mobius, which gives
 \bea
 {\cal A}^{(k)}_{\mc{N}=1} &\xrightarrow[l' \rightarrow \infty]{UV}& -\left( \frac{B}{4\pi^2} \right)^2  \int_{l'}^{\infty}  dl\; \hbox{Tr}\left[ \half \left( q_1^2\gamma_{k} \otimes \gamma^{-1}_{k} + 2q_1\gamma_{k} \otimes \gamma^{-1}_{k}q_2 + \gamma_{k} \otimes \gamma^{-1}_{k} q_2^2\right) \right]  \prod_{i=1}^3  \left| 2 \sin \pi \theta^k_i \right|   \;, \nonumber \\
{\cal M}_{\mc{N}=1}^{(k)}  &\xrightarrow[l' \rightarrow \infty]{UV}& -\left( \frac{B}{4\pi^2} \right)^2  \int_{l'}^{\infty} dl \;\hbox{Tr} \left[ 4 q^2 \gamma_{\Omega'_{k}} \gamma^{-T}_{\Omega'_{k}} \right] \prod_{i=1}^3 s_i \left( - 2 \sin\left(\pi R_i^k\right) \right) \;. \label{muv1bexp}
\eea
The $\mc{N}=4$ untwisted sectors do not contribute to the $B^2$ terms due to supersymmetry.
There are also $\mc{N}=2$ contributions and these take the exact form
\bea
{\cal A}_{\mc{N}=2}^{(k)} &=&   \left( \frac{B}{4\pi^2} \right)^2  \int_0^{\infty} \frac{dt}{2t}  \hbox{Tr} \left[  \left(q^2_1 \gamma_{k} \otimes \gamma^{-1}_{k} + 2q_1\gamma_{k} \otimes \gamma^{-1}_{k}q_2 + \gamma_{k} \otimes \gamma^{-1}_{k} q^2_2 \right)   \right]
\cos\left(\pi\theta_3^k\right)\prod^{2}_{i=1} \left(2\sin \left( \pi \theta^k_i \right) \right) \;,  \nonumber \\
{\cal M}_{\mc{N}=2}^{(k)} &=&   \left( \frac{B}{4\pi^2} \right)^2  \int_0^{\infty} \frac{dt}{2t}  \hbox{Tr} \left[  -2q^2 \gamma_{\Omega'_{k}} \gamma^{-T}_{\Omega'_{k}}  \right]
\cos\left(\pi R_3^k\right)\prod^{2}_{i=1} \left(2\sin \left( \pi R^k_i \right) \right) \;,
\eea
where the product is over the two twisted angles and and $\theta_3$ (and $R_3$) denote the untwisted direction.
These expressions are exact due to the $\mc{N}=2$ structure. With $\mc{N}=2$ supersymmetry only
BPS multiplets can renormalise the gauge couplings and the string oscillator tower is all non-BPS.
As a result in a purely local computation the only non-zero contribution comes from the zero modes.

 Evaluated in the IR limit $t \to \infty$
 the magnetised $B^2$ amplitudes must reproduce the field theory $\beta$-functions.
 Evaluated in the UV limit $t \to 0$ the amplitudes give the threshold corrections to the gauge couplings. As discussed in the introduction the key feature of the $\mc{N}=2$ sector is that, since the expressions are exact, the running is with the same coefficient in the IR and the UV.
Evaluated in a purely local model, such $\mc{N}=2$ sectors give logarithmic ultraviolet divergences, $\int \frac{dt}{t} \Delta^{\mc{N}=2}$.
This ultraviolet divergence is associated with a tadpole for partially twisted field.
In a global model
these divergences are cutoff as global tadpole cancellation occurs. From the closed string channel, this corresponds to the existence
of new brane/O-plane sectors located in the bulk which also act as sources for the partially twisted field and cancel the tadpole
sourced in the local model.

From the open string viewpoint the incorporation of these sectors corresponds to
the inclusion of winding modes from the singularity to the distant bulk branes/O-planes.
Such modes are charged and BPS and contribute to the threshold corrections,
cutting off the $\beta$-function running from the $\mc{N}=2$ sector.
The details of the cutoff depend on the precise and model-dependent location of the bulk branes, but what is
model-independent is that the winding modes act as an effective ultraviolet cutoff on the $\mc{N}=2$ sector,
cutting off the running at a mass scale $M_X = R M_s$.

For the $\mc{N}=1$ sectors there is no such decoupling. In the IR the $\mc{N}=1$ sector
combines with the $\mc{N}=2$ contributions to give the field theory $\beta$ functions.
In the UV the string oscillator tower enters giving a non-vanishing contribution.
For non-abelian generators the threshold corrections vanish in the far UV
as closed string tadpole cancellation is enforced.
For $U(1)$ generators the threshold corrections can diverge due to an on-shell exchange of a $\mc{N}=1$ twisted RR mode via a Green-Schwarz coupling $C_2 \wedge \mathrm{Tr} F$. The abelian case is discussed in detail in appendix \ref{sec:anu1} but will not feature in the main text.

Similar to the way winding modes give an effective cutoff at $M_X = R M_s$ for $\mc{N}=2$ sectors, the oscillator modes give an
effective cutoff at $M_s$ for $\mc{N}=1$ sectors. As $\mc{N}=1$ sectors are purely local tadpole cancellation occurs once the open string
 oscillators are included. As $t \ll 1$ gives $l \sim 1/t \gg 1$, in this limit all higher closed string modes are
 exponentially decoupled and the amplitude
 reduces to the (vanishing) IR closed string tadpole. Modulo small corrections that do not depend on the overall volume, the effective
 cutoff for the $\mc{N}=1$ sectors is therefore at $t = 1/M_s^2$.

The general amplitude therefore looks like (\ref{stringthres}) which we recall here
\be
\frac{1}{g^2}(\mu) = \frac{1}{g^2} \Big\vert_0 + \beta_a \ln \left( \frac{M_s^2}{\mu^2} \right) +
\beta_a^{\mc{N}=2} \ln \left( \frac{M_X^2}{M_s^2} \right).
\ee
As this involves different running between $M_X$ and $M_s$ and $M_s$ and $\mu$ depending on the relative size of
$\mc{N}=1$ and $\mc{N}=2$ contributions to the beta function, in general this appears to differ with the Kaplunovsky-Louis formula
(\ref{KL}) which only contains field theory running from the winding scale $M_X$.

In the next sections we shall study this issue in detail for the $\mbb{C}^3/\mbb{Z}_4$ orientifold, and see how the
discrepancy can be resolved.

\subsection{The $\mathbb{Z}_4$ case}
\label{sec:stringthresz4}

We now specialise the above formulae to the case of the $\mathbb{Z}_4$ orientifold. We begin by turning on a background field within the $SU(n_1)$ gauge group.
Note that since the orientifold identifies $SU(n_1)$ and $SU(n_3)$ we must turn on the background field for both. The normalisation is fixed by the canonical gauge field normalisation $\mathrm{Tr} q^2 = \half$.
 The charge matrices are then given by
\bea
q = q_1 = -q_2 &=& \mathrm{diag}\left( {\bf 0}_{n_0}, Q_{SU(n_1)}, {\bf 0}_{n_2}, Q_{SU(n_1)}\right) \;, \\
Q_{SU(n_1)} &=& \frac{1}{\sqrt{8}} \mathrm{diag} \left( 1, -1, 0, ..., 0\right) \;.
\eea

Evaluating the amplitudes from section \ref{sec:magamp}, summing over the $\mc{N}=1$ and $\mc{N}=2$ sectors separately we find explicitly
\bea
{\cal A}^{SU(n_1)}_{\mc{N}=1} &=& \frac{1}{4} \sum_{k=1,3} {\cal A}_{\mc{N}=1}^{(k)} = 0 \;, \nonumber \\
{\cal M}^{SU(n_1)}_{\mc{N}=1} &=& \frac{1}{4} \sum_{k=1,2,3,4} {\cal M}_{\mc{N}=1}^{(k)} = 0 \;, \\
{\cal A}^{SU(n_1)}_{\mc{N}=2} &=& \frac{1}{4} {\cal A}_{\mc{N}=2}^{(2)} =  -\left( 2n_1 - 2n_2 - 4\right) \left( \frac{B}{4\pi^2} \right)^2  \int_0^{\infty} \frac{dt}{2t}    \;, \nonumber \\
{\cal M}^{SU(n_1)}_{\mc{N}=2} &\equiv& 0 \;.
\eea
We can extract the $\beta_{SU(n_1)}$ function by imposing IR and UV cutoffs on the $\mc{N}=2$ integral
set by the probe energy scale $1/\mu^2$ and the winding modes scale $M_X = R M_s$ respectively. Then using (\ref{vactoamp}) we get
\bea
\frac{\Lambda_2^{SU(n_1)}}{8\pi^2} & = & -\frac{1}{8\pi^2} \left( 2n_1 - 2n_2 - 4 \right) \int_{\frac{1}{M^2_{X}}}^{\frac{1}{\mu^2}} \frac{dt}{2t} \nonumber \\
& = &  \beta_{SU(n_1)} \mathrm{ln} \left( \frac{M_X^2}{\mu^2} \right)\;.
\eea
which exactly matches the expected field theory result using (\ref{1loopvac}) and (\ref{z4sun1beta}). This is the same behaviour that was observed in \cite{Conlon:2009xf} for the purely orbifold case.

We now turn to the $SO(n_0)$ gauge group and turn on the generator
\bea
q = q_1 = -q_2 &=& \mathrm{diag}\left( Q_{SO(n_0)}, {\bf 0}_{n_1}, {\bf 0}_{n_2}, {\bf 0}_{n_1} \right) \;, \\
Q_{SO(n_0)} &=& \frac{i}{2} \left( \begin{array}{cccc} 0 & 1 & 0 & ...  \\ -1 & 0 & 0 & ...  \\ 0 & 0 & 0 & ...  \\ ... & ... & ... & ...   \end{array} \right) \;.
\eea
Evaluating the amplitudes we find, using the tadpoles (\ref{z4tadpole}),
\bea
{\cal A}^{SO(n_0)}_{\mc{N}=1} &=& \frac{1}{4} \sum_{k=1,3} {\cal A}_{\mc{N}=1}^{(k)} \xrightarrow[t' \rightarrow \infty]{IR} -\left( \frac{B}{4\pi^2} \right)^2  \int_{t'}^{\infty}  \frac{dt}{2t} 2 \left( n_0 - n_2\right)  = +8 \left( \frac{B}{4\pi^2} \right)^2  \int_{t'}^{\infty}  \frac{dt}{2t}  \;, \nonumber \\
{\cal M}^{SO(n_0)}_{\mc{N}=1} &=& \frac{1}{4} \sum_{k=1,2,3} {\cal M}_{\mc{N}=1}^{(k)} \xrightarrow[t' \rightarrow \infty]{IR} 6 \left( \frac{B}{4\pi^2} \right)^2  \int_{t'}^{\infty}  \frac{dt}{2t}   \;, \\
{\cal A}^{SO(n_0)}_{\mc{N}=2} &=& \frac{1}{4} {\cal A}_{\mc{N}=2}^{(2)} =  -\left( n_0 + n_2 - 2n_1 \right) \left( \frac{B}{4\pi^2} \right)^2  \int_0^{\infty} \frac{dt}{2t}  =   -\left( 2n_0 - 2n_1 + 4\right) \left( \frac{B}{4\pi^2} \right)^2  \int_0^{\infty} \frac{dt}{2t}    \;, \nonumber \\
{\cal M}^{SO(n_0)}_{\mc{N}=2} &\equiv& 0 \;.
\eea
As described above tadpole cancellation ensures that in the UV the $\mc{N}=1$ annulus and Mobius
amplitudes cancel against each other.
We can check this UV cancellation using the expressions (\ref{muv1bexp}) which give
\bea
\label{uvv1}
{\cal A}^{SO(n_0)}_{\mc{N}=1} &\xrightarrow[l' \rightarrow \infty]{UV}& -\left( \frac{B}{4\pi^2} \right)^2  \int_{l'}^{\infty} dl \;4\left( n_0 - n_2 \right) =  \left( \frac{B}{4\pi^2} \right)^2  \int_{l'}^{\infty} dl \;16 \;, \\
{\cal M}_{\mc{N}=1}^{SO(n_0)}  &\xrightarrow[l' \rightarrow \infty]{UV}& -\left( \frac{B}{4\pi^2} \right)^2  \int_{l'}^{\infty} dl \;16  \;,
\label{uvv2}
\eea
leading to $\Lambda_2^{SO(n_0)} \xrightarrow{UV} 0$. From closed string channel the non-cancelling
subleading terms in (\ref{uvv1}) and (\ref{uvv2}) are of order $e^{- \pi l/M_s^2}$ and so vanish exponentially
once $l \gg M_s^2$ or $t \ll 1/M_s^2$.
Therefore, up to small additional corrections we obtain an
effective cutoff at $M_s$ for the $\mc{N}=1$ amplitudes and
an effective cutoff at $M_X$ for $\mc{N}=2$ amplitudes.
To compare with the Kaplunovsky-Louis expression
we impose these cutoffs on the $\mc{N}=1$ and $\mc{N}=2$ sector, also taking the IR cutoff at $1/\mu^2$.
We find
\bea
\frac{\Lambda_2^{SO(n_0)}}{8\pi^2} &=& \frac{1}{16\pi^2} \left(-14 \right)\; \mathrm{ln} \left( \frac{\mu^2}{M_s^2} \right) + \frac{1}{16\pi^2}  \left( 2n_0 - 2n_1 + 4 \right)  \mathrm{ln} \left( \frac{\mu^2}{M_X^2} \right) \nonumber \\
 &=&  \frac{1}{16\pi^2} \left(2n_0 - 2n_1 - 10 \right) \mathrm{ln} \left( \frac{\mu^2}{M_s^2} \right) +  \frac{1}{16\pi^2}  \left( 2n_0 - 2n_1 + 4 \right)  \mathrm{ln} \left( \frac{M_s^2}{M_X^2} \right) \nonumber \\
 &\equiv& \left( \beta_{SO(n_0)}^{\mc{N}=1} +  \beta_{SO(n_0)}^{\mc{N}=2} \right)  \mathrm{ln} \left( \frac{M_s^2}{\mu^2} \right) + \beta_{SO(n_0)}^{\mc{N}=2} \mathrm{ln} \left( \frac{M_X^2}{M_s^2} \right)  \label{z4betan1n2} \\
 &=& \beta_{SO(n_0)}\mathrm{ln} \left( \frac{M_s^2}{\mu^2} \right) + \beta_{SO(n_0)}^{\mc{N}=2} \mathrm{ln} \left( \frac{M_X^2}{M_s^2} \right) \;,
\eea
where we defined $\beta_{SO(n_0)}^{\mc{N}=1}$ and $\beta_{SO(n_0)}^{\mc{N}=2}$ as the contributions in the IR to the $\beta$ functions coming from the $\mc{N}=1$ and $\mc{N}=2$ sectors respectively. The expression (\ref{z4betan1n2}) differs from the naive application of the KL formula (\ref{mirage}). The difference arising from a non-vanishing contribution to the $\beta$ function from the $\mc{N}=1$ sector.

To study the $Sp(n_2)$ gauge group we turn on the generator
\bea
q = q_1 = -q_2 &=& \mathrm{diag}\left({\bf 0}_{n_0}, {\bf 0}_{n_1},  Q_{Sp(n_2)}, {\bf 0}_{n_1} \right) \;, \\
Q_{Sp(n_2)} &=& \frac{1}{2} \left( \begin{array}{cccc} 1 & 0 & 0 & ...  \\ 0 & -1 & 0 & ...  \\ 0 & 0 & 0 & ...  \\ ... & ... & ... & ...   \end{array} \right) \;.
\eea
In a similar fashion this gives the result
\be
\frac{\Lambda_2^{Sp(n_2)}}{8\pi^2} = \left( \beta_{Sp(n_2)}^{\mc{N}=1} +  \beta_{Sp(n_2)}^{\mc{N}=2} \right)  \mathrm{ln} \left( \frac{M_s^2}{\mu^2} \right) + \beta_{Sp(n_2)}^{\mc{N}=2} \mathrm{ln} \left( \frac{M_X^2}{M_s^2} \right) \;,
\ee
where
\bea
\beta_{Sp(n_2)}^{\mc{N}=1} &=&  \frac{1}{16\pi^2} \left(14\right) = - \beta_{SO(n_0)}^{\mc{N}=1} \;, \\
\beta_{Sp(n_2)}^{\mc{N}=2}  &=&  \frac{1}{16\pi^2}  \left( 2n_0 -2n_1 + 4 \right) = -\beta_{SO(n_0)}^{\mc{N}=2} \;.
\eea
The physics is therefore the same as the $SO(n_0)$ case: the effective $\beta$ function undergoes a jump at $M_s$ and so there are
two distinct phases of the running, one from $M_X$ to $M_s$ and one from $M_s$ to $\mu$.

Gathering these results together, we have
\bea
\frac{\Lambda_2^{SO(n_0)}}{8\pi^2} & = &  \left( \Delta +  \beta_{SO(n_0)}^{\mc{N}=2} \right)  \mathrm{ln} \left( \frac{M_s^2}{\mu^2} \right) + \beta_{SO(n_0)}^{\mc{N}=2} \mathrm{ln} \left( \frac{M_X^2}{M_s^2} \right) \;, \nonumber \\
\frac{\Lambda_2^{SU(n_1)}}{8\pi^2} & = & \beta^{\mc{N}=2}_{SU(n_1)} \mathrm{ln} \left( \frac{M_s^2}{\mu^2} \right) +
\beta^{\mc{N}=2}_{SU(n_1)} \mathrm{ln} \left( \frac{M_X^2}{M_s^2} \right)\;, \nonumber \\
\frac{\Lambda_2^{Sp(n_2)}}{8\pi^2} & = & \left( -\Delta +  \beta_{Sp(n_2)}^{\mc{N}=2} \right)  \mathrm{ln} \left( \frac{M_s^2}{\mu^2} \right) + \beta_{Sp(n_2)}^{\mc{N}=2} \mathrm{ln} \left( \frac{M_X^2}{M_s^2} \right) \;,
\label{delt}
\eea
where $\Delta = \beta_{SO(n_0)}^{\mc{N}=1} = -\beta_{Sp(n_2)}^{\mc{N}=1}$.

The form of the gauge couplings differ from the naive application of the KL formula as in (\ref{mirage}).
The difference lies in the presence of the $\Delta$ term in equations (\ref{delt}) associated to $\mc{N}=1$ sectors
which contribute to running from the string scale but not from the winding scale.
We now proceed to study how this discrepancy is resolved.

\section{Threshold corrections: matching the field theory}
\label{sec:matchfield}

Recall that the string calculation gives the coefficient multiplying the $F^2$ term in the Lagrangian which in the field theory also includes a tree-level coupling to chiral superfields $M_k$
\be
\label{gcs}
{\cal L} = \left( \frac{1}{g_a^2} +  \sum_{i=1}^{K} s^a_{k} \mathrm{Re}{M}_k \right) \mathrm{Tr} F_a^2 \;.
\ee
The coupling (\ref{gcs}) comes from the holomorphic gauge kinetic function
\be
f_a(\Phi) = S + s^a_{k} M_k \;.
\ee
The chiral superfields $M_k$ correspond to closed string twisted modes with $\mathrm{Re}M_k$ corresponding to the NS-NS part and $\mathrm{Im}M_k$ the RR part.

Geometrically the twisted modes correspond to collapsed two- and four-cycles. They can therefore be thought of as dimensionally reducing the NS and RR supergravity form fields on the collapsed two-cycle.
These fields descend from
reducing either $J \wedge J + iC_4$ or $B_2 + i C_2$ on collapsed
two/four-cycles.\footnote{The NS two-form $B_2$ splits into a part that is even and a part that is odd under the orientifold action $B_2=B_2^+ + B_2^-$. $B_2^-$ parameterises the modulus (by abuse of notation we label this $B_2$ in the main text) and therefore vanishes (at tree level) at the singularity. The field $B_2^+$ can have a non-vanishing vev at the singularity as in \cite{Bachas:2008jv, Aspinwall:1995zi}.}
Here $J$ is the Kahler form, $B_2$ the NS two-form, and $C_2$ and $C_4$ are the RR two and four-form respectively.
Depending on whether we reduce the above fields on cycles or their dual cycles we can obtain either linear or chiral multiplets depending on
whether the bosonic 4d fields associated to the reduction of $C_4$ are scalars or 2-forms. We denote the chiral multiplet by $M_k$ and the real scalar component of the linear multiplet by $m_k$.

In \cite{9402005} the supergravity analysis that led to the KL formula (\ref{KL}) is carried out using
chiral multiplets.
Therefore to compare the string result with the supergravity formula we need to dualise the linear multiplet to a chiral multiplet. This procedure is described in detail in \cite{Derendinger:1991hq} for the heterotic string. In appendix D we review this and also discuss the IIB case.
 Performing a similar analysis for the local IIB models the result is that at tree level\footnote{The analysis is basically the same as the heterotic case \cite{Derendinger:1991hq} but with the dilaton replaced by the twisted mode. The only significant change is that the Kahler potential for the twisted mode is quadratic $K \sim m_k^2$ rather than logarithmic.}
\be
\hbox{Re} M_k = m_k \;,
\ee
with $m_k$ the linear multiplet and $M_k$ the chiral multiplet. $m_k$ always vanishes at the singularity where $M_k$ only vanishes
at tree-level.
 However at 1-loop level this is modified to (\ref{redef}) which we reproduce here
\be
\hbox{Re}(M_k) = m_k - \alpha_k \ln R^2 \;. \label{redef2}
\ee
An important point is that the 1-loop field redefinition (\ref{redef2}) is tied to the 1-loop correction to the gauge kinetic function and is therefore only present when the particular $\mc{N}=1$ twisted mode $m_k$ contributes to the $\beta$ functions. As discussed in the introduction, this precisely reproduces the behaviour required to match the string and field theory results if the coupling $s^a_k$ are proportional to the $\beta^{(k)}_a$.
In the next section we explicitly perform the string calculation to check this proportionality.

\subsection{Extracting closed string couplings}

\begin{figure}[ht]
\label{1loop}
\begin{center}
\includegraphics[width=12cm]{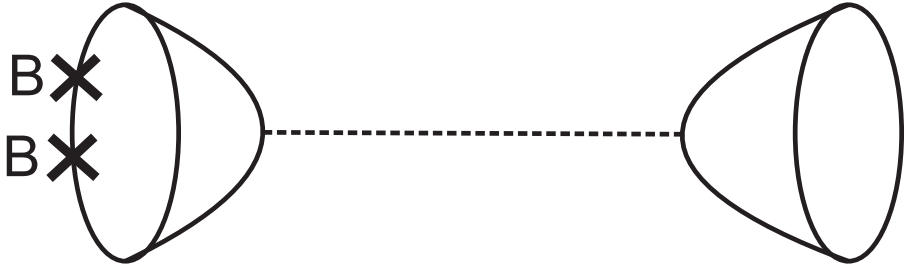}
\caption{From a supergravity perspective this represents an $\left(\hbox{Re}M_k\right) F_{\mu \nu} F^{\mu \nu}$ vertex sourcing an $M_k$ field
which propagates and is then absorbed by the vacuum tadpole. By factoring out the vacuum tadpole we can infer the
coefficient of the $\left(\hbox{Re}M_k\right) F_{\mu \nu} F^{\mu \nu}$ coupling.}
\end{center}
\end{figure}

In order to calculate the correction to the gauge couplings
induced by the field redefinitions we need to know the
coupling $s^a_{k}$ of the twisted closed string
modes to the gauge field strengths $F^a_{\mu \nu} F^{a,\mu \nu}$. One way to calculate this
is by extracting the UV divergence of the $B^4$ amplitude which corresponds to
exchanging an on-shell twisted modes sourced by the magnetic field background \cite{Antoniadis:1999ge}.
However this method only gives $\left(s_{ak}\right)^2$ and so is insensitive to the sign of $s_{ak}$ which for us plays a crucial role. The method we employ is to study the open string UV divergence of the $B^2$ Annulus amplitudes. In the closed string tree level channel this amplitude can be interpreted as a vertex between the NSNS closed string twisted mode and the gauge field strength, sourcing the twisted mode which is then absorbed by the vacuum, see figure \ref{1loop}.
Of course the overall diagram vanishes once tadpole cancellation is imposed but the $\left(\hbox{Re}M_k\right) F_{\mu \nu}F^{\mu \nu}$ coupling can be extracted by stripping off the tadpole piece which just corresponds to the trace over the other end of the string. This is because the tadpoles give this coupling for the RR fields but due to supersymmetry these are equivalent up to a constant to the NS tadpoles.\footnote{Each twisted sector gives rise to two real closed string modes and their coupling is given by the real and imaginary parts of $\mathrm{Tr}\gamma^k_{\theta}$. However in our orientifolds the imaginary part will always vanish corresponding to projecting out that twisted mode.}

\subsection{The $\mathbb{Z}_4$ case}

In this section we calculate the $\mc{N}=1$ twisted mode coupling to the gauge fields for the $\mathbb{Z}_4$ case and show that it takes the form appropriate for reconciling the string calculation of section \ref{sec:threstr} with the field theory KL formula (\ref{KL}).

The Annulus UV amplitudes read
\bea
{\cal A}^{SO(n_0)}_{\mc{N}=1} &\xrightarrow[l' \rightarrow \infty]{UV}& -\left( \frac{B}{4\pi^2} \right)^2  \int_{l'}^{\infty} dl 4 \left( n_0 - n_2 \right) \;,\\
{\cal A}^{Sp(n_2)}_{\mc{N}=1} &\xrightarrow[l' \rightarrow \infty]{UV}& -\left( \frac{B}{4\pi^2} \right)^2  \int_{l'}^{\infty} dl 4 \left( n_2 - n_0 \right) \;,\\
{\cal A}^{SU(n_1)}_{\mc{N}=1} &\xrightarrow[l' \rightarrow \infty]{UV}& 0 \;.
\eea
There is a single closed string twisted mode $m_0$ and its coupling to the vacuum is given by $\mathrm{Tr}\gamma_{\theta}=\left(n_0-n_2\right)$ which gives
\be
s^{SO(n_0)}_0 = \frac{\beta_{SO(n_0)}^{\mc{N}=1}}{\alpha_0} \;,\; s^{Sp(n_2)}_0 = \frac{\beta_{Sp(n_2)}^{\mc{N}=1}}{\alpha_0}  \;,\; s^{SU(n_1)}_0 = 0 \;.
\ee
Here $\alpha_0$ is some (gauge group) universal constant that corresponds to extracting the appropriately normalised coupling and propagator. We therefore find the required result that the coupling are proportional to the $\mc{N}=1$ $\beta$ functions. Recall that in terms of the gauge kinetic functions this reads
\bea
f_{SO(n_0)} & = & S +  \frac{\beta_{SO(n_0)}^{\mc{N}=1}}{\alpha_0} M_0 \;, \\
f_{SU(n_1)} & = & S \;, \\
f_{Sp(n_2)} & = & S + \frac{\beta_{Sp(n_2)}^{\mc{N}=1}}{\alpha_0}M_0.
\eea

We therefore see that if at the orbifold point the chiral superfield
\be
\mathrm{Re}M_0= -\alpha_0 \mathrm{ln\;R^2}   \;,
\ee
the holomorphic gauge couplings become non-universal.
The string results then match exactly the field theory formula with
\be
s^a_0 \mathrm{Re}M_0 = -\beta_a^{\mc{N}=1} \mathrm{ln}\left( \frac{M_X^2}{M_s^2}\right) \;.
\ee
The striking point is that a single field redefinition is capable of altering three $\beta$ functions in a way to resolve the discrepancy
with the naive use of the Kaplunovsky-Louis formula.

\section{More examples}
\label{sec:moreexa}

In this section we present two more examples of orientifolded singularities, $\mbb{Z}_6$ and $\mbb{Z}_6^{'}$, that
serve as checks on the above
analysis and understanding. As with the $\mathbb{Z}_4$ case, as far as we are aware these
have not been previously presented in the literature and so we outline their construction before moving on to the magnetised amplitude calculations.
These orientifolds exhibits more structure compared to the $\mathbb{Z}_4$ example through the presence of more $\mc{N}=1$/$\mc{N}=2$ twisted closed string modes.

\subsection{The $\mbb{C}^3/\mathbb{Z}_6$ orientifold}

The $\mathbb{Z}_6$ orbifold action is generated by $\theta=\left(1/6,1/6,-1/3\right)$. We take the orientifold spatial action to be $R = \left(7/12,1/12,-2/3 \right)$. The orientifold group is therefore
\bea
& &\left\{ \left(0, 0, 0 \right), \left(\frac16, \frac16, -\frac13 \right),  \left(\frac13, \frac13, -\frac23 \right),  \left(\frac12, \frac12, -1 \right), \left(\frac23, \frac23, -\frac43 \right), \left(\frac56, \frac56, -\frac53 \right),\right.  \nonumber  \\
& & \;\; \Omega I \left(\frac{7}{12}, \frac{1}{12}, -\frac23 \right), \Omega I \left(\frac34, \frac14, -1 \right), \Omega I \left(\frac{11}{12}, \frac{5}{12}, -\frac43 \right), \Omega I \left(\frac{13}{12}, \frac{7}{12}, -\frac53 \right), \nonumber \\
& & \;\;\left.   \Omega I \left( \frac54, \frac34, -2 \right), \Omega I \left(\frac{17}{12}, \frac{11}{12}, -\frac{7}{3} \right)  \right\},
\eea
Including the spatial action of $I$ the fixed point locus consists solely of the origin.
We take the orbifold generating element
\be
\gamma_{\theta} =  \mathrm{diag}(\unit_{n_0}, \alpha \unit_{n_1}, \alpha^2 \unit_{n_2}, \alpha^3 \unit_{n_3}, \alpha^4 \unit_{n_4}, \alpha^5 \unit_{n_5}) \quad {\rm with}\;\; \alpha=e^{\pi i/3} \;,
\ee
and impose $n_1=n_5$ and $n_2=n_4$. For the orientifold action we take
\be
\gamma_{\Omega'}  = \left( \begin{array}{cccccc}
\unit_{n_0} & 0 & 0 &  0 & 0 & 0 \\
0 & 0 & 0 & 0 & 0 &  \alpha^{1/2}\epsilon_{n_5}  \\
0 & 0 & 0 & 0 & \alpha \epsilon_{n_4} & 0 \\
0 & 0 &0 &  \alpha^{3/2}\epsilon_{n_3} & 0 & 0 \\
0 & 0 & \alpha^2 \epsilon_{n_2}  & 0 & 0 & 0\\
0 & \alpha^{5/2} \epsilon_{n_1}  & 0 &  0 & 0 & 0
 \end{array}\right) \;,
\ee
with $\epsilon$ denoting the anti-symmetric matrix with unit off-diagonal entries.

Calculating the tadpoles using (\ref{tad1}-\ref{tad2}) leads to
\be
\hbox{Tr}\left[\gamma_{\theta}\right] - 8 = \hbox{Tr}\left[\gamma_{\theta^2}\right]  = 0 \;. \label{z6gammatad}
\ee
There are two real closed string $\mc{N}=1$ twisted modes modes $m_0$ and $m_1$ associated with the first and second tadpole conditions in (\ref{z6gammatad}) respectively.
The tadpole constraints (\ref{z6gammatad}) impose the conditions
\bea
n_2 &=& n_0 - 4 \;, \nonumber \\ \label{z6tadpoles}
n_3 &=& n_1 - 4 \;.
\eea

The massless fermionic spectrum of the theory can be calculated using (\ref{massfer}) which gives the matter content shown in table 2. The gauge group is
\be
G = SO(n_0) \times U(n_1) \times U(n_2) \times Sp(n_3) \;.
\ee
\begin{table}
\center
\label{tab:z6orientifold}
\begin{tabular}{|c|c|c|c|c|}
\hline
Multiplicity & \multicolumn{4}{|c|}{Representation}  \\
\hline
& $SO(n_0)$ & $SU(n_1)$ & $SU(n_2)$ & $Sp(n_3)$ \\
\hline
2 & $n_0$ & $\bar{n}_1$ & 1           & 1 \\
2 & 1     & $n_1$       & $\bar{n}_2$ & 1 \\
2 & 1     &   1         & $n_2$       & $n_3$\\
1 & $n_0$ &  1          & $n_2$       & 1 \\
1 & 1     & $\bar{n}_1$ & 1           & $n_3$ \\
1 & 1     & 1           & $\bar{A}_{n_2}$  & 1 \\
1 & 1     & $S_{n_1}$   & 1           & 1 \\
\hline
\end{tabular}
\caption{Field content and representations for $Z_6$ orientifold. The $n_i$ denote the fundamental representation and $S$ and $A$ denote symmetric and anti-symmetric representations respectively.}
\end{table}
The non-abelian anomalies of the theory correspond to (\ref{z6tadpoles}). We are also interested in the beta functions for the gauge groups and these read, after imposing anomaly cancellation (\ref{z6tadpoles}),
\bea
\beta_{SO(n_0)} &=& \frac{1}{16\pi^2}\left( -2n_0 + 2n_1 + 2 \right) = -\beta_{Sp(n_3)}\;,  \\
\beta_{SU(n_1)} &=& \frac{1}{16\pi^2}\left( 2n_0 - 2n_1 - 5\right) = -\beta_{SU(n_2)}\;.
\eea

The calculation for the threshold corrections proceeds as in section \ref{sec:stringthresz4} and here we quote the relevant results. The vacuum energies read
\be
\frac{\Lambda_2^{a}}{8\pi^2} =  \left( \beta_{a}^{\mc{N}=1} +  \beta_{a}^{\mc{N}=2} \right)  \mathrm{ln} \left( \frac{M_s^2}{\mu^2} \right) + \beta_{a}^{\mc{N}=2} \mathrm{ln} \left( \frac{M_X^2}{M_s^2} \right)   \;,
\ee
where we have
\bea
 \beta_{SO(n_0)}^{\mc{N}=1} &=& -\beta_{Sp(n_3)}^{\mc{N}=1} = \frac{1}{16\pi^2}\left( -2 \right) \;,\;\; \beta_{SO(n_0)}^{\mc{N}=2} = -\beta_{Sp(n_3)}^{\mc{N}=2} = \frac{1}{16\pi^2}\left(  -2n_0 + 2n_1 + 4 \right)\;, \\
 \beta_{SU(n_1)}^{\mc{N}=1} &=&  -\beta_{SU(n_2)}^{\mc{N}=1} = -\frac{1}{16\pi^2} \;,\;\; \beta_{SU(n_1)}^{\mc{N}=2} = - \beta_{SU(n_2)}^{\mc{N}=2} = \frac{1}{16\pi^2}\left( 2n_0 - 2n_1 - 4\right)\;.
\eea
An important point is that the $\mc{N}=1$ contributions to $\Lambda_2^{a}$ comes solely from the $\gamma_{\theta}$ sector. In terms of the $\mc{N}=1$ twisted modes $m_0$ and $m_1$ we have (as in section \ref{sec:intsummary})
\bea
\beta_a^0 &=& \beta_a^{\mc{N}=1} \;, \nonumber \\
\beta_a^1 &=& 0 \;. \label{betatwistz6}
\eea

The gauge couplings take the form
\bea
\frac{\Lambda_2^{SO(n_0)}}{8\pi^2} & = & \left( - 2 \Delta +  \beta_{SO(n_0)}^{\mc{N}=2} \right)  \mathrm{ln} \left( \frac{M_s^2}{\mu^2} \right) + \beta_{SO(n_0)}^{\mc{N}=2} \mathrm{ln} \left( \frac{M_X^2}{M_s^2} \right)   \;, \nonumber \\
\frac{\Lambda_2^{SU(n_1)}}{8\pi^2} & = & \left( -  \Delta +  \beta_{SU(n_1)}^{\mc{N}=2} \right)  \mathrm{ln} \left( \frac{M_s^2}{\mu^2} \right) + \beta_{SU(n_1)}^{\mc{N}=2} \mathrm{ln} \left( \frac{M_X^2}{M_s^2} \right) \;, \nonumber \\
\frac{\Lambda_2^{SU(n_2)}}{8\pi^2} & = & \left( \Delta +  \beta_{SU(n_2)}^{\mc{N}=2} \right)  \mathrm{ln} \left( \frac{M_s^2}{\mu^2} \right) + \beta_{SU(n_2)}^{\mc{N}=2} \mathrm{ln} \left( \frac{M_X^2}{M_s^2} \right)  \;, \nonumber \\
\frac{\Lambda_2^{Sp(n_3)}}{8\pi^2} & = & \left( 2 \Delta +  \beta_{Sp(n_3)}^{\mc{N}=2} \right)  \mathrm{ln} \left( \frac{M_s^2}{\mu^2} \right) + \beta_{Sp(n_3)}^{\mc{N}=2} \mathrm{ln} \left( \frac{M_X^2}{M_s^2} \right)  \label{z6betan1n2} \;,
\eea
where as before $\Delta = \beta_{SO(n_0)}^{\mc{N}=2}$.

We can now check that the closed string twisted modes couple in the correct way to match the string calculation with the field theory results. In this case we have two closed string twisted modes $m_0$ and $m_1$. We find
\bea
s^{SO(n_0)}_0 &=& - s^{Sp(n_3)}_0 =  \frac{\beta_{SO(n_0)}^{\mc{N}=1}}{\alpha_0}  \;,\;\;  s^{SO(n_0)}_1 = s^{Sp(n_3)}_1 =  \frac{\beta_{SO(n_0)}^{\mc{N}=1}}{\alpha_1}  \;, \nonumber \\
s^{SU(n_1)}_0 &=& - s^{SU(n_2)}_0 =  \frac{\beta_{SU(n_1)}^{\mc{N}=1}}{\alpha_0}  \;,\;\;  s^{SU(n_1)}_1 = s^{SU(n_2)}_1 =  -\frac{\beta_{SU(n_1)}^{\mc{N}=1}}{\alpha_1}  \;.
\eea
Here $\alpha_0$ and $\alpha_1$ are constants (different from their $\mathbb{Z}_4$ values). In terms of the gauge kinetic functions this reads
\bea
f_{SO(n_0)} & = & S + \frac{\beta_{SO(n_0)}^{\mc{N}=1}}{\alpha_0} M_0 + \frac{\beta_{SO(n_0)}^{\mc{N}=1}}{\alpha_1}  M_1\;, \nonumber \\
f_{SU(n_1)} & = & S + \frac{\beta_{SU(n_1)}^{\mc{N}=1}}{\alpha_0} M_0 - \frac{\beta_{SU(n_1)}^{\mc{N}=1}}{\alpha_1}  M_1\;, \nonumber \\
f_{SU(n_2)} & = & S + \frac{\beta_{SU(n_2)}^{\mc{N}=1}}{\alpha_0} M_0 + \frac{\beta_{SU(n_2)}^{\mc{N}=1}}{\alpha_1}  M_1\;, \nonumber \\
f_{Sp(n_3)} & = & S +  \frac{\beta_{Sp(n_3)}^{\mc{N}=1}}{\alpha_0} M_0 -  \frac{\beta_{Sp(n_3)}^{\mc{N}=1}}{\alpha_1} M_1 \;.
\eea
There is new structure compared to the $\mathbb{Z}_4$ case. We now have two linear multiplet $\mc{N}=1$ twisted modes $m_0$ and $m_1$.
$m_0$ has an appropriate coupling proportional to the $\beta$ functions, but $m_1$ does not as its
coupling to the $SU(n_1)$ and $Sp(n_3)$ gauge groups has the wrong sign. Therefore we expect that  in this case $M_0$ undergoes a one-loop redefinition restoring consistency with
the Kaplunovsky-Louis formula while $M_1$ does not undergo a redefinition and still has vanishing vev at the singularity. This matches the fact that the 1-loop redefinition is proportional to the contribution to the $\beta$ functions and the result (\ref{betatwistz6}).

From the string perspective the redefinition is related to the fact that fractional O-planes wrap the $m_1$ collapsed cycle and contribute to the $m_1$ tadpole, while O-planes
do not wrap the $m_2$ cycle and in closed string channel this tadpole is wholly sourced by annulus diagrams.

\subsection{The $\mbb{C}^3/\mathbb{Z}'_6$ orientifold}

The $\mathbb{Z}'_6$ orbifold action is generated by $\theta=\left(1/6,1/3,-1/2\right)$. We take the orientifold spatial action to be $R = \left(7/12,-1/3,-1/4 \right)$. The orientifold group is therefore
\bea
& &\left\{ \left(0, 0, 0 \right), \left(\frac16, \frac13, -\frac12 \right),  \left(\frac13, \frac23, -1\right),  \left(\frac12, 1, -\frac32 \right), \left(\frac23, \frac43, -2 \right), \left(\frac56, \frac53, -\frac52 \right),\right.  \nonumber  \\
& & \;\; \Omega I \left(\frac{7}{12}, -\frac13, -\frac14 \right), \Omega I \left(\frac34, 0, -\frac34 \right), \Omega I \left(\frac{11}{12}, \frac13, -\frac54 \right), \Omega I \left(\frac{13}{12}, \frac{2}{3}, -\frac74 \right), \nonumber \\
& & \;\;\left.   \Omega I \left( \frac54, 1, -\frac94 \right), \Omega I \left(\frac{17}{12}, \frac43, -\frac{11}{4} \right)  \right\},
\eea
The tadpoles are
\be
\hbox{Tr}\left[\gamma_{\theta}\right] + 4 =  0 \;. \label{zp6gammatad}
\ee
This is associated with a single $\mc{N}=1$ twisted closed string mode.
The CP embedding is the same as for $\mathbb{Z}_6$ but now the tadpole constraint reads
\be
-n_0 - n_1 + n_2 + n_3 - 4 = 0 \;. \label{z6ptadpoles}
\ee
The massless fermionic spectrum gives the matter content shown in table 3 and the gauge group is the same as $\mathbb{Z}_6$.
\begin{table}
\center
\label{tab:z6porientifold}
\begin{tabular}{|c|c|c|c|c|}
\hline
Multiplicity & \multicolumn{4}{|c|}{Representation}  \\
\hline
& $SO(n_0)$ & $SU(n_1)$ & $SU(n_2)$ & $Sp(n_3)$ \\
\hline
1 & $n_0$ & $\bar{n}_1$ & 1           & 1  \\
1 & 1     & $n_1$       & $\bar{n}_2$ & 1  \\
1 & 1     & 1           & $n_2$       & $n_3$  \\
1 & $n_0$ & 1           & $\bar{n}_2$ & 1  \\
1 & 1     & $n_1$       & 1           & $n_3$  \\
1 & $n_0$ & 1           & 1           & $n_3$  \\
1 & 1     & $n_1$       & $n_2$       & 1  \\
1 & 1     & $\bar{n}_1$ & $\bar{n}_2$  & 1  \\
1 & 1     & 1           & $A_{n_2}$   & 1  \\
1 & 1     & $\bar{S}_{n_1}$  & 1 & 1 \\
\hline
\end{tabular}
\caption{Field content and representations for $\mathbb{Z}'_6$ orientifold. The $n_i$ denote the fundamental representation and $S$ and $A$ denote symmetric and anti-symmetric representations respectively.}
\end{table}
The non-abelian anomalies match the tadpoles (\ref{z6ptadpoles}). Using the tadpoles (\ref{z6ptadpoles}) to eliminate $n_3$, the $\beta$ functions read
\bea
\beta_{SO(n_0)} &=& \frac{1}{16\pi^2}\left( -2n_0 + 2n_1 + 10   \right) \;,  \\
\beta_{SU(n_1)} &=& \frac{1}{16\pi^2}\left( n_0 - 2n_1 + n_2 + 3   \right) \;, \\
\beta_{SU(n_2)} &=& \frac{1}{16\pi^2}\left( n_0 + 2n_1 -3n_2 + 1    \right) \;,  \\
\beta_{Sp(n_3)} &=&  \frac{1}{16\pi^2}\left( -2n_0 - 2n_1 + 4n_2 - 18 \right) \;.
\eea

The calculation for the threshold corrections gives the vacuum energies
\be
\frac{\Lambda_2^{a}}{8\pi^2} =  \left( \beta_{a}^{\mc{N}=1} +  \beta_{a}^{\mc{N}=2} \right)  \mathrm{ln} \left( \frac{M_s^2}{\mu^2} \right) + \beta_{a}^{\mc{N}=2} \mathrm{ln} \left( \frac{M_X^2}{M_s^2} \right)  \;,
\ee
where we have
\bea
\beta_{SO(n_0)}^{\mc{N}=1} &=& -\beta_{Sp(n_3)}^{\mc{N}=1} = \frac{1}{16\pi^2}\left( 10 \right) \;,\\
\beta_{SU(n_1)}^{\mc{N}=1} &=&  -\beta_{SU(n_2)}^{\mc{N}=1} = \frac{1}{16\pi^2} \left( 5 \right) \;, \\
\beta_{SO(n_0)}^{\mc{N}=2} &=& \frac{1}{16\pi^2}\left( -2n_0 + 2n_1 \right) \;, \\
\beta_{SU(n_1)}^{\mc{N}=2} &=& \frac{1}{16\pi^2}\left( n_0 - 2n_1 + n_2 - 2\right) \;, \\
\beta_{SU(n_2)}^{\mc{N}=2} &=& \frac{1}{16\pi^2}\left( n_0 + 2n_1 - 3n_2 + 6\right) \;, \\
\beta_{Sp(n_3)}^{\mc{N}=2} &=&  \frac{1}{16\pi^2}\left( -2n_0 - 2n_1 + 4n_2 - 8 \right)\;.
\eea
We can now check that the closed string twisted modes couple in the correct way to match the string calculation with the field theory results. In this case we have one closed string twisted mode $m_0$. We find
\bea
s^{SO(n_0)}_0 &=& - s^{Sp(n_3)}_0 = \frac{\beta_{SO(n_0)}^{\mc{N}=1}}{\alpha_0} \;, \nonumber \\
s^{SU(n_1)}_0 &=& - s^{SU(n_2)}_0 = \frac{\beta_{SU(n_1)}^{\mc{N}=1}}{\alpha_0} \;.
\eea
We see again that the closed string mode $m_0$ has the appropriate coupling to match the field theory results after the appropriate redefinition.

\subsection{D3-D7 Orbifolds}

We can also resolve here a puzzle encountered in \cite{Conlon:2009xf}. In that paper
systems of D3/D7 branes on the $\mbb{C}^3/\mbb{Z}_3$ orbifold singularity were considered.
As described in \cite{aiqu} such systems
give phenomenologically promising spectra. It was found that the string computation for threshold
corrections to the D3 gauge couplings gave
universal running between the string and winding scale and non-universal running below the string scale.

We can resolve this issue and find that full agreement with Kaplunovsky-Louis can be found through a
redefinition of the two twisted moduli. We briefly summarise the results but for full details of the models
refer to \cite{aiqu, Conlon:2009xf}. The holomorphic gauge couplings are
\bea
f_{SU(n_0)} & = & S + M_1, \nonumber \\
f_{SU(n_1)} & = & S - \frac{M_1}{2} + \frac{\sqrt{3}}{2} M_2, \nonumber \\
f_{SU(n_2)} & = & S - \frac{M_1}{2} - \frac{\sqrt{3}}{2} M_2.
\eea
The $\beta$ functions can be written as
\be
\frac{\Lambda_2^{a}}{8\pi^2} =  \left( \beta_{a}^{\mc{N}=1} +  \beta_{a}^{\mc{N}=2} \right)  \mathrm{ln} \left( \frac{M_s^2}{\mu^2} \right) + \beta_{a}^{\mc{N}=2} \mathrm{ln} \left( \frac{M_X^2}{M_s^2} \right) \;,
\ee
with
\bea
\beta_{SU(n_0)}^{\mc{N}=1} &=& \frac{1}{16\pi^2} \left( \frac{2 n_0^7 - n_1^7 - n_2^7}{3} \right), \nonumber \\
\beta_{SU(n_1)}^{\mc{N}=1} &=& \frac{1}{16\pi^2} \left(\frac{- n_0^7 + 2 n_1^7 - n_2^7}{3} \right), \nonumber \\
\beta_{SU(n_2)}^{\mc{N}=1} &=& \frac{1}{16\pi^2} \left(\frac{- n_0^7 - n_1^7 + 2 n_2^7}{3} \right).
\eea
Through the redefinitions
\bea
\hbox{Re} M_1 &=& m_1 + \frac{1}{16\pi^2}  \left( \frac{2 n_0^7 - n_1^7 - n_2^7}{3} \right) \ln R^2, \\
\frac{\sqrt{3}}{2} \hbox{Re} M_2 &=& \frac{\sqrt{3}}{2} m_2 +  \frac{1}{16\pi^2} \left( \frac{n_1^7 - n_2^7}{2} \right) \ln R^2,
\eea
we can obtain a full match with Kaplunovsky-Louis. In this case the non-trivial aspect of the redefinition is that
two field redefinitions are sufficient to match three gauge couplings. Note that in this case there are two
fields $M_1$ and $M_2$ contributing from the same twisted sector which did not occur in the orientifold models. This slightly modifies the scenario so that the relationship between the twisted mode gauge coupling $s_a^k$ and the $\beta_a$ functions is generalised.

\section{Conclusions}

In this work we studied threshold corrections to the gauge couplings in local models of branes at orientifold singularities.
This extends the work of \cite{Conlon:2009xf} on threshold corrections at orbifold singularities and provides new tractable examples
of local models with full CFT control. For local models
the general supergravity analysis performed by Kaplunovsky and Louis \cite{9303040, 9402005}
points towards a unification scale that is enhanced by the bulk radius from the string scale, with field theory running below
this winding mode scale.

Our aim has been to understand this formula and address the question of when running starts at the winding mode scale and when running
starts at the string scale.
We analysed this issue using explicit string calculations and showed that in general this apparent unification at an enhanced scale is only present in particular constructions. The low-energy gauge couplings take the form
\be
\frac{1}{g^2}(\mu) = \frac{1}{g^2} \Big\vert_0 + \beta_a \ln \left( \frac{M_s^2}{\mu^2} \right) +
\beta_a^{\mc{N}=2} \ln \left( \frac{M_X^2}{M_s^2} \right),
\ee
where $M_X$ is the winding mode scale.
Unification of the gauge coupling at the enhanced scale $M_X$ occurs whenever there are no non-universal $\mc{N}=1$ fully twisted, that is modes confined to the singularity, contributions to the gauge coupling threshold corrections. Such modes give contributions to gauge coupling running
that start at $M_s$.
If there is no $\mc{N}=1$ contribution then the apparent unification at the enhanced scale can be understood \cite{Conlon:2009xf} from the fact that the remaining $\mc{N}=2$ contribution gives field theory running up to the winding mode scale where cancellation of global tadpoles implies that the winding modes cut the running off.

We performed detailed calculations to show how this can be reconciled with the Kaplunovsky-Louis formula by an appropriate field redefinition of the $\mc{N}=1$ closed string modes. This arises from dualising the string linear multiplets to the supergravity chiral multiplets used in the Kaplunovsky-Louis analysis. This required particular couplings of these modes to the gauge fields and the string calculations showed that these couplings are indeed as required.

Understanding the form of $\mc{N}=1$ and $\mc{N}=2$ contributions
to the gauge coupling threshold corrections is therefore important in understanding gauge coupling unification in local models.
 Open-closed string duality relates the threshold corrections to tadpoles and
 for orbifold/orientifold models
 gives a relatively simple rule as to when this occurs: if there is a contribution from multiple diagrams (that transform differently from the IR to the UV)  to the local tadpoles then $\mc{N}=1$ sectors contribute threshold corrections to the gauge couplings. In the examples we presented the D3-D3 cylinder was supplemented by contributions from the Mobius strip in case of orientifolds or the D3-D7 cylinder in the case of D7 branes present.

The fact that generic branes at singularities local models only exhibit gauge coupling unification up to corrections logarithmic in the bulk radius can be attributed to the fact that they are not GUT models: the different gauge groups originate from different branes and the
tree-level unification of gauge couplings at the singularity is accidental and does not survive at 1-loop.
This comes from the fact that in general twisted sectors
 couple non-universally to the gauge groups.
Although naively it appears that holomorphic gauge couplings are universal at the singularity, we have seen that at one loop level
this is not the case, and the non-GUT nature of the setup becomes apparent.

  This implies that for true local GUTs this will not occur. We will discuss threshold corrections in this case
  in \cite{ToAppear}. It would also be interesting to study the mirror type IIA picture with intersecting D6 branes. There the Kaplunovsky-Louis formula would again imply that there can be a significant gap between the apparent unification scale and the string scale, especially within the weakly-coupled models of \cite{08041248}. However there is no geometric picture of a local model and so it would be interesting to understand the relevant string physics. Finally it is clear that questions regarding gauge coupling in local models can only be fully answered once the dynamics of the blow-up fields are understood.

\subsection*{Acknowledgments}
We thank Mark Goodsell, Andre Lukas, Fernando Quevedo, Christoffer Petersson, Erik Plauschinn and Graham Ross for useful discussions and explanations. JC is supported by a Royal Society University Research Fellowship. EP is supported by a STFC Postdoctoral Fellowship.
JC thanks Cooks Branch and CERN for hospitality during the course of the work.

\newpage
\appendix

\section{Anomalous $U(1)$s}
\label{sec:anu1}

In this section we study anomalous $U(1)$s in branes at singularity models. More precisely we study the Green-Schwarz mechanism for generating $U(1)$ masses, which automatically operates in the case of anomalous $U(1)$s but can also affect non-anomalous $U(1)$s. This topic does not quite fall within the narrative of the main text and is therefore relegated to the appendix. Further, the cancellation of $U(1)$ anomalies for models of branes at singularities has been studied following the initial work of \cite{Ibanez:1998qp}. However, the physics of anomalous $U(1)$s has some overlap with the gauge threshold corrections discussed in the main text. We also present our analysis using the background field method which differs from the analysis of \cite{Ibanez:1998qp}. We therefore present this appendix as containing partially new results but primarily to illuminate the physics associated to $U(1)$s in local models.

The physics of interest is of course the Green-Schwarz mechanism. The relevant terms in the four dimensional theory are $C_0 F \wedge F$ and $C_2 \wedge F$.\footnote{It is possible to think of these as geometrically arising from say a D7 brane with the Chern-Simons term reduced on a collapsing four-cycle $\Sigma$ with two-cycle submanifolds $\omega_{\alpha}$. We can write
\be
\int_{D7} C_4 \wedge F \wedge F =  \int_{\Sigma} C_4 \int_{M_4} F \wedge F + \int_{\Sigma} \omega_{\alpha} \wedge f \int_{M_4} C_2^{\alpha} \wedge F \;,
\ee
where $f$ denotes the world volume flux. A key point here is that the two-cycles $\omega_\alpha$ need not be globally homologically non-trivial, in which case there is no propagating field associated with $C_2^{\alpha}$ and no Green-Schwarz coupling. As this property requires a global completion
to determine, this corresponds to the $\mc{N}=2$ twisted sector fields being non-normalisable in the non-compact geometry as discussed in this appendix. At the CFT level this corresponds to a logarithmic divergence, which may either vanish or be enhanced to a quadratic divergence in the
presence of the global completion.}
We can extract these terms in the action using the background field method. The (open string UV, closed string IR) divergence for the $B^2$ and $B^4$ amplitudes correspond to the on-shell exchange of the RR mode $C_2$ and the NS partner of the $C_0$ field respectively. By turning on a background field for the most general combination of $U(1)$s we can extract which $U(1)$ combinations have each type of coupling. This determines which $U(1)$s participate in Green-Schwarz anomally canellation and/or become massive.

For purposes of simplicity and clarity we perform the calculations within a $\mathbb{Z}_4$ orbifold setting (with no orientifolds).
The $\mathbb{Z}_4$ orbifold action is generated by $\theta=\left(1/4,1/4,-1/2\right)$. Since there are no orientifolds the tadpoles are sourced purely by the Annulus diagram and therefore read
\be
\hbox{Tr}\left[\gamma_{\theta}\right]  = 0 \;.
\ee
We take the orbifold generating element
\be
\gamma_{\theta} =  \mathrm{diag}(\unit_{n_0}, \alpha \unit_{n_1}, \alpha^2 \unit_{n_2}, \alpha^3 \unit_{n_3}) \quad {\rm with}\;\; \alpha=e^{\pi i/2} \;,
\ee
The tadpole constraints impose the condition
\bea
n_0 &=& n_2  \;, \nonumber \\
n_1 &=& n_3 \;. \label{z4orbtadpole}
\eea
The massless fermionic spectrum of the theory can be calculated using (\ref{orbspect}) which gives the matter content shown in table 4.
\begin{table}
\center
\label{tab:z4orbifold}
\begin{tabular}{|c|c|}
\hline
Multiplicity & Representation  \\
\hline
2 & $\left(n_0, \bar{n}_1\right)$ \\
2 & $\left(n_1, \bar{n}_2\right)$ \\
2 & $\left(n_2, \bar{n}_3\right)$ \\
2 & $\left(n_3, \bar{n}_0\right)$ \\
1 & $\left(n_0, \bar{n}_2\right)$ \\
1 & $\left(n_2, \bar{n}_0\right)$ \\
1 & $\left(n_1, \bar{n}_3\right)$ \\
1 & $\left(n_3, \bar{n}_1\right)$ \\
\hline
\end{tabular}
\caption{Field content and representations for $Z_4$ orbifold. The bracket pairs are bi-fundamental representations.}
\end{table}
The non-abelian anomalies of the theory correspond to (\ref{z4orbtadpole}). We are also interested in the abelian and mixed anomalies. We consider a general $U(1)$ combination
\be
U(1)_Y \equiv Y_0 U(1)_0 +  Y_1 U(1)_1 +  Y_2 U(1)_2 +  Y_3 U(1)_3 \;.
\ee
Then, after imposing the tadpoles, the mixed and abelian gauge anomalies are given by
\bea
{\cal A}_{SU(n_0)^2-U(1)_Y} &=& -{\cal A}_{SU(n_2)^2-U(1)_Y}  = 2 n_1 (Y_3 - Y_1) \nonumber \\
{\cal A}_{SU(n_1)^2 - U(1)_Y} &=& - {\cal A}_{SU(n_3)^2 - U(1)_Y}  =  2 n_0 (Y_0 - Y_2) \nonumber \\
{\cal A}_{U(1)_Y^3} & = & -6 n_0 n_1 \left[ (Y_1 - Y_3)(Y_0^2 - Y_2^2) - (Y_0 - Y_2)(Y_1^2 - Y_3^2) \right]. \label{z4orbanomal}
\eea

To extract the relevant coupling we turn on the background field
\be
q_1 = -q_2 = \frac{1}{\mc{N}} \left( Y_0 \unit_{n_0}, \, Y_1 \unit_{n_1}, Y_2 \unit_{n_2}, Y_3 \unit_{n_3} \right) \;,
\ee
where $\mc{N} = \sqrt{2\sum_{a=0}^3 Y^2_a n_a}$.

\subsubsection*{The $\mc{N}=1$ sector}

The $B^2$ coefficient for the Annulus amplitude in the UV (\ref{auvbexpapp}) for the $\mc{N}=1$ sector can be decomposed as
\be
{\cal A}_{\mc{N}=1}^{B^2} = {\cal A}_{\mc{N}=1}^{q_1^2 + q_2^2} + {\cal A}_{\mc{N}=1}^{q_1 q_2} \;.
\ee
The amplitude ${\cal A}_{\mc{N}=1}^{q_1^2 + q_2^2 }$ corresponds to the tadpoles exactly as in the case of the non-abelian generators studied in the main sections. The amplitude ${\cal A}_{\mc{N}=1}^{q_1 q_2}$ vanished for the non-abelian case because the generators were traceless. However it is non vanishing for the abelian case and in the UV gives the divergence associated with a $C_2 \wedge F$ coupling in the action with $C_2$ being an $\mc{N}=1$ twisted RR mode in this case. In the closed string channel the amplitude reads
\be
{\cal A}_{\mc{N}=1}^{q_1 q_2} \xrightarrow[l' \rightarrow \infty]{} \left( \frac{B}{4\pi^2} \right)^2  \int_{l'}^{\infty}  dl \frac{8}{{\cal N}^2} \left[ n^2_0 \left( Y_0 - Y_2  \right)^2 + n^2_1 \left( Y_1 - Y_3  \right)^2 \right]  \;.
\ee
We see that this gives precisely the coupling needed to cancel mixed anomalies, and half of the appropriate expressions for the cubic abelian anomalies (\ref{z4orbanomal}). Any $U(1)_Q$ for which this amplitude is non-vanishing gains a mass. The $B^4$ amplitude gives the coupling $C_0 F \wedge F$ and reads
\be
{\cal A}_{\mc{N}=1}^{q^2_1 q^2_2} \xrightarrow[l' \rightarrow \infty]{} -B^2 \left( \frac{B}{4\pi^2} \right)^2  \int_{l'}^{\infty}  dl \frac{2}{{\cal N}^4} \left[ n^2_0 \left( Y^2_0 - Y^2_2  \right)^2 + n^2_1 \left( Y^2_1 - Y^2_3  \right)^2 \right]  \;.
\ee
This takes the form required to match the expression for the cubic anomalies. We have therefore checked that the $\mc{N}=1$ RR field couples in the correct way to cancel the abelian anomalies.

\subsubsection*{The $\mc{N}=2$ sector}

The closed string $\mc{N}=2$ sector is sensitive to the global geometry.
Extracting the coupling by studying the UV divergence of the $B^2$ and $B^4$ amplitudes is more complicated since it depends on the global completion of the local model. This is reflected in terms of winding modes affecting the amplitude above the winding scale. Indeed the local, in the sense of not including winding modes, amplitudes take the form
\bea
{\cal A}_{\mc{N}=2}^{q_1 q_2} &\xrightarrow[l' \rightarrow \infty]{}& \left( \frac{B}{4\pi^2} \right)^2  \int_{l'}^{\infty}  \frac{dl}{l} \frac{8}{{\cal N}^2} \left[ n_0 \left( Y_0 + Y_2  \right) - n_1 \left( Y_1 + Y_3  \right) \right]^2   \;,\\
{\cal A}_{\mc{N}=2}^{q^2_1 q^2_2} &\xrightarrow[l' \rightarrow \infty]{}& -B^2 \left( \frac{B}{4\pi^2} \right)^2  \int_{l'}^{\infty}  \frac{dl}{l} \frac{1}{{\cal N}^4} \left[ n_0 \left( Y^2_0 + Y^2_2  \right) - n_1 \left( Y^2_1 + Y^2_3  \right) \right]^2 \;.
\eea
The $\mc{N}=2$ divergence is logarithmic rather than linear, as was the case for the $\mc{N}=1$ sector, which reflects the fact that the closed string $\mc{N}=2$ modes are not propagating physical modes without a global completion. This implies that they cannot participate in the anomaly cancellation since this is a local property. However they can still induce a Green-Schwarz mass for $U(1)$ fields depending on the global completion. Indeed we see that the dependence of the amplitudes on the $Y_a$ is not directly related to the field theory anomalies.

\begin{figure}[ht]
\label{z4orbifold}
\begin{center}
\includegraphics[width=12cm]{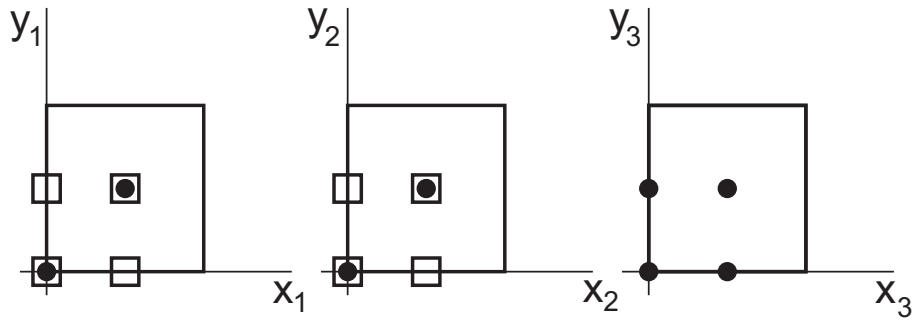}
\caption{The $T^6/\mbb{Z}_4$ orbifold. Dark circles correspond to $\theta$ fixed points and hollow squares correspond to
$\theta^2$ fixed points.}
\end{center}
\end{figure}

We now want to investigate the physics once we compactify the space. As our testbed we will use the $T^6/\mbb{Z}_4$ orbifold
shown in figure \ref{z4orbifold}. We will introduce an additional brane stack to cancel twisted tadpoles. We will not cancel untwisted tadpoles;
while clearly this is necessary in a full compactification it does not affect the physics of interest here.
As a compact space this orbifold has $h^{1,1} = 31, h^{2,1} =7$. The 31 elements of $h^{1,1}$ decomposes as
5 untwisted 2-cycles, 16 $\theta^1$ twisted cycles stuck at the 16 $\mbb{Z}_4$ fixed points, 6 $\theta^2$ twisted cycles stuck
at $\mbb{Z}_4$ invariant combinations of $\theta^2$ fixed points, and 4 $\theta^2$ twisted cycles at $\mbb{Z}_4$ fixed points and
propagating across the third $T^2$.

We place a single stack of fractional branes at the origin $(0,0,0)$ (point A) of multiplicity $(n_0, n_1, n_2, n_3) = (N, M, N, M)$.
As in \cite{Conlon:2009xf} we also introduce a stack of fractional branes on the (0,0,i/2) (point B) of multiplicity
$(n_0, n_1, n_2, n_3) = (M, N, M, N)$. This cancels the $\mc{N}=2$ twisted tadpoles.
The effect of the compact space and additional stack of branes is to modify the $q_1q_2$ amplitude.
We must now include the $AA$ winding modes which give an extra factor
$\sum_{n,m} e^{- \pi R^2 t (n^2 + m^2)}$.\footnote{There is also potentially an $AB$ winding mode stack with a factor
$\sum_{n,m} e^{- \pi R^2 t ((n+1/2)^2 +m^2)}$, which does not contribute to the $q_1 q_2$ sector.} In the UV limit $l \to \infty$ this gives
\bea
\sum_{n,m} e^{- \pi R^2 t (n^2 + m^2)} & \to & \frac{2l}{R^2} \left( 1 + \mc{O}(e^{- \frac{2 \pi l}{R^2}}) \right),
\eea
The amplitude therefore becomes
\be
{\cal A}_{\mc{N}=2}^{q_1 q_2} \xrightarrow[l' \rightarrow \infty]{} \left( \frac{B}{4\pi^2} \right)^2  \int_{l'}^{\infty}  \frac{dl}{l} \frac{2l}{R^2} \frac{8}{{\cal N}^2} \left[ n_0 \left( Y_0 + Y_2  \right) - n_1 \left( Y_1 + Y_3  \right) \right]^2   \;.
\ee
This is now linearly divergent and corresponds to a physical coupling
\be
\frac{m_s}{R} \left(n_0 (Y_0 + Y_2) - n_1 (Y_1 + Y_3) \right) C_2 \wedge F \;, \label{n2mass}
\ee
which induces a mass for the field.

We can now determine the masses of the non-anomalous $U(1)$s in the model. There are two orthogonal non-anomalous $U(1)$s to consider
\bea
U(1)_{diag} &=& \frac{1}{n_0}(U(1)_0 + U(1)_2) + \frac{1}{n_1}(U(1)_1 + U(1)_3) \;, \\
U(1)_{tw} &=& \frac{1}{n_1}(U(1)_0 + U(1)_2) - \frac{1}{n_0}(U(1)_1 + U(1)_3) \;.
\eea
From (\ref{n2mass}) we see that $U(1)_{diag}$ remains massless while $U(1)_{tw}$ gains a mass.

Actually within a global context $C_2$ can couple to both the $A$ and $B$ stacks of branes.
We therefore ought to consider a general $U(1)$ which combines both stacks of branes,
$U(1) = \sum_a Y_a U(1)_a^A + Z_a U(1)_a^B$. As neither $U(1)^A_{diag}$ nor $U(1)^B_{diag}$ have any coupling to the
twisted sectors, the same is true of linear combinations of these, which therefore remain massless in the compact model.
The interesting case, which we focus on, is the combination $U(1)_{tw}^A \pm U(1)_{tw}^B$. We can in fact verify that for
\be
U(1)_X = \left( \frac{1}{n_1}(U(1)_0 + U(1)_2) - \frac{1}{n_0}(U(1)_1 + U(1)_3) \right) -
\left( \frac{1}{n_0}(U(1)_0 + U(1)_2) - \frac{1}{n_1}(U(1)_1 + U(1)_3) \right) \;,
\ee
then ${\cal A}_{\mc{N}=2}^{q_1q_2} \rightarrow 0$. This implies that the amplitude has no quadratic
divergence and $U(1)_X$ has no $C_2 \wedge F$ coupling to give it a mass. In contrast $U(1)_{tw}^A + U(1)_{tw}^B$ has a nonvanishing
${\cal A}_{\mc{N}=2}^{q_1q_2}$ and becomes massive with a mass given by $\sim m_s/R$ consistent with
\cite{0610007, 08105660}.

In total there are then three massless orthogonal $U(1)$s present:
\bea
U(1)^A_{diag} &=& \frac{1}{n_0}(U(1)^A_0 + U(1)^A_2) + \frac{1}{n_1} (U(1)^A_1 + U(1)^A_3) \;,  \\
U(1)^B_{diag} &=& \frac{1}{n_1}(U(1)^B_0 + U(1)^B_2) + \frac{1}{n_0} (U(1)^B_1 + U(1)^B_3) \;, \nonumber \\
U(1)_X &=& \left( \frac{1}{n_1}(U(1)^A_0 + U(1)^A_2) - \frac{1}{n_0} (U(1)^A_1 + U(1)^A_3) \right) \nonumber \\
& & 
- \left( \frac{1}{n_0}(U(1)^B_0 + U(1)^B_2) - \frac{1}{n_1} (U(1)^B_1 + U(1)^B_3) \right) \;. \nonumber
\eea
The third $U(1)$ mixes the `visible' and `hidden' sector. The details of this $U(1)$, and the fields that are charged under it,
depend on the precise nature of the hidden sector.

This shows explicitly at the CFT level how the masses of non-anomalous $U(1)$s are determined depending on the global geometry.
Analysed locally, we obtain a logarithmic divergence. If the cycle is globally trivial, then as we extend to a fully global model
the divergence vanishes and the Green-Schwarz coupling is absent. If the cycle is non-trivial, then the logarithmic divergence
becomes a quadratic divergence suppressed by the bulk radius. This quadratic divergence signals the presence of the Green-Schwarz term and
the $U(1)$ mass.

The same techniques using the background field formalism should be applicable for the related problem of studying kinetic mixing among separate
$U(1)$s which may be interesting to carry out. This would be complementary to the vertex operator techniques used in
\cite{08031449}.

\section{Tadpoles}
\label{sec:tadapp}

In this appendix we calculate the tadpole divergences for local orientifolds.  We are interested in $\mc{N}=1$ fully twisted tadpoles that receive contributions form the annulus, Mobius strip, and Klein bottle one-loop amplitudes. The annulus amplitude is given by (we work in units with $2\alpha'=1$)
\be
{\cal A} = -\int_{0}^{\infty} \frac{dt}{2t} \mathrm{STr}\left[ \frac{1}{N} \sum_{k=0}^N \theta^k \left( \frac{1+\left(-1\right)^F}{2} \right) q^{\left(p_\mu p^{\mu} + M^2\right)/2} \right] \;.
\ee
Here $t$ parameterises the annulus width with $t \rightarrow 0$ corresponding to the UV and $t \rightarrow \infty$ the IR. The sum over $\theta^k$ imposes the orbifold projection and $\left( \frac{1 + (-1)^F}{2} \right)$ the GSO projection.  $q=e^{2\pi i \tau} = e^{-\pi t}$ where $\tau$ is the torus parameter for the annulus $\tau=it/2$. $p_{\mu}$ and $M$ are the momentum and mass of the string states. The $\mathrm{STr}$ stands for tracing over the bosons and fermions as $\sum_{\mathrm{NS}} - \sum_{\mathrm{RR}}$ . Performing this trace we can write
\be
{\cal A} =  \frac{1}{N} \sum_{k=0}^N {\cal A}^{(k)} \;,
\ee
with
\be
{\cal A}_{\mc{N}=1}^{(k)} =  - \int \frac{dt}{2t} \frac{1}{(2 \pi^2 t)^2} \hbox{Tr} \left[ \gamma_k \otimes \gamma_k^{-1} \right]
\sum_{\alpha, \beta=0,1/2}  \frac{\eta_{\alpha \beta}}{2}  \frac{\vartheta \Big[ \begin{array}{c} \alpha \\ \beta \end{array} \Big]}{\eta^3}
 \prod_{i=1}^3 \left( - 2 \sin \pi \theta^k_i \right) \frac{\vartheta \Big[ \begin{array}{c} \alpha \\ \beta + \theta^k_i \end{array} \Big]}{\vartheta \Big[ \begin{array}{c} 1/2 \\ 1/2+\theta^k_i \end{array} \Big]} \;.
\ee
Here the $\eta$ and $\theta$ functions are functions of $q$ and are explicitly given in appendix \ref{confor}. We have $\eta_{\alpha\beta}=\left(-1\right)^{2\left(\alpha + \beta - 2 \alpha \beta\right)}$. The trace is coming from the trace over the CP indices. The terms with $\alpha=0$  and $\alpha=1/2$ are due to RR and NS states respectively.

Similarly we have for the Mobius strip and Klein bottle
\be
{\cal M}_{\mc{N}=1}^{(k)} =   \int \frac{dt}{2t} \frac{1}{(2 \pi^2 t)^2} \hbox{Tr} \left[ \gamma_{\Omega'_{k}} \gamma_{\Omega'^{-T}_{k}} \right]
\sum_{\alpha, \beta=0,1/2}  \frac{\eta_{\alpha \beta}}{2}  \frac{\vartheta \Big[ \begin{array}{c} \alpha \\ \beta \end{array} \Big]}{\eta^3}
 \prod_{i=1}^3 \left( - 2 \sin\left(\pi R_i^k\right) \right) \frac{\vartheta \Big[ \begin{array}{c} \alpha \\ \beta + R_i^k \end{array} \Big]}{\vartheta \Big[ \begin{array}{c} 1/2 \\ 1/2 +R_i^k \end{array} \Big]} \;,
\ee
\be
{\cal K}_{0}^{(k)} =   4\int \frac{dt}{2t} \frac{1}{(2 \pi^2 t)^2}
\sum_{\alpha, \beta=0,1/2}  \frac{\eta_{\alpha \beta}}{2} \frac{\vartheta \Big[ \begin{array}{c} \alpha \\ \beta \end{array} \Big]}{\eta^3}
\prod_{i=1}^3 \left(\frac{ - 2 \sin \left(2\pi R_i^k \right) }{ 4 \cos^2 \left( \pi R_i^k \right) }\right) \frac{\vartheta \Big[ \begin{array}{c} \alpha \\ \beta + 2R_i^k \end{array} \Big]}{\vartheta \Big[ \begin{array}{c} 1/2 \\ 1/2+2R_i^k \end{array} \Big]} \;,
\ee
\be
{\cal K}_{2}^{(k)} =   4\int \frac{dt}{2t} \frac{1}{(2 \pi^2 t)^2}
\sum_{\alpha, \beta=0,1/2}  \frac{\eta_{\alpha \beta}}{2} \frac{\vartheta \Big[ \begin{array}{c} \alpha \\ \beta \end{array} \Big]}{\eta^3}
\prod_{i=1}^3 \left( \frac{ - 2 \sin \left(2\pi R_i^k \right)}{ 4 \cos^2 \left(\pi R_i^k \right)}\right)^{\delta_i} \frac{\vartheta \Big[ \begin{array}{c} \alpha + \theta_i^{N/2} \\ \beta + 2R_i^k \end{array} \Big]}{\vartheta \Big[ \begin{array}{c} 1/2 +   \theta_i^{N/2} \\ 1/2+2R_i^k \end{array} \Big]} \;.
\ee
Here
\be
R_i^k = \theta^k_i + R_i \;.
\ee
The last amplitude ${\cal K}_{2}^{(k)}$ only occurs for even orientifolds and in that case
\be
\delta_i = \left\{ \begin{array}{l} 0\;\mathrm{if}\; \;\;\theta_i^{N/2} \;\mathrm{mod}\; 1 = 1/2 \\  1 \;\mathrm{otherwise}  \end{array} \right. \;.
\ee
It corresponds to the contribution for the $Z_2$ twisted closed string sector with all other contributions, apart from the untwisted sector ${\cal K}_{0}^{(k)}$, vanishing by symmetry. The Mobius strip amplitude is a function of $-q$ and the Klein bottle of $q^2$ corresponding to torus parameters of $\tau=1/2+it/2$ and $\tau=it$ respectively. The extra factors of $\cos^2$ in the denominators are due to zero mode integration over the internal space \cite{GimonJohnson}.

To extract the UV divergence we can modular transform, using formulae in appendix E, to the (tree level) closed string channel through the transformations $t=1/l$, $t=1/4l$, $t=1/2l$ for the annulus, Mobius strip and Klein bottle respectively. $l$ is now the cylinder length and the UV limit is given by $l \rightarrow \infty$. Performing the transformation we find\footnote{Since the Mobius amplitude is a function of $-q$ the transformation needs to be done through a series of transformations of the torus parameter $1/2 + it/2 = \tau \rightarrow -\frac{1}{\tau} \rightarrow  -\frac{1}{\tau} + 2 \rightarrow \left( \frac{1}{\tau} - 2 \right)^{-1} = 2il - 1/2$ \cite{Antoniadis:1999ge}.}
\be
{\cal A}_{\mc{N}=1}^{(k)} =  \frac{i}{4}  \int \frac{dl}{\left(2\pi^2\right)^2} \hbox{Tr} \left[  \gamma_k \otimes \gamma_k^{-1} \right]
\sum_{\alpha, \beta=0,1/2} \frac{\eta_{\alpha \beta}}{2}  \frac{\vartheta \Big[ \begin{array}{c} \beta \\ -\alpha \end{array} \Big]}{\eta^3}
 \prod_{i=1}^3 \left( - 2 \sin \pi \theta_i \right) \frac{\vartheta \Big[ \begin{array}{c} \beta + \theta_i \\ -\alpha \end{array} \Big]}{\vartheta \Big[ \begin{array}{c} 1/2 + \theta_i\\ -1/2 \end{array} \Big]}
\ee
\be
{\cal M}_{\mc{N}=1}^{(k)} =  -2i \int \frac{dl}{\left(2\pi^2\right)^2} \hbox{Tr} \left[ \gamma_{\Omega'_{k}} \gamma_{\Omega'^{-T}_{k}} \right]
\sum_{\alpha, \beta=0,1/2} \frac{\eta_{\alpha \beta}}{2}  \frac{\vartheta \Big[ \begin{array}{c} \alpha \\ \beta \end{array} \Big]}{\eta^3}
 \prod_{i=1}^3 \left( - 2 \sin \left(\pi R_i^k\right) \right) \frac{\vartheta \Big[ \begin{array}{c} \alpha+ 2R_i^k  \\ \beta + R_i^k \end{array} \Big]}{\vartheta \Big[ \begin{array}{c} 1/2  + 2R_i^k \\ 1/2 + R_i^k \end{array} \Big]}  \;.
\ee
\be
{\cal K}_{0}^{(k)} =  -4i \int \frac{dl}{\left(2\pi^2\right)^2}
\sum_{\alpha, \beta=0,1/2} \frac{\eta_{\alpha \beta}}{2}  \frac{\vartheta \Big[ \begin{array}{c} \beta \\ -\alpha \end{array} \Big]}{\eta^3}
\frac{\prod_{i=1}^3 \left( - 2 \sin 2\pi R_i^k \right)}{\prod_{i=1}^3 4 \cos^2 \left( \pi R_i^k \right) }  \frac{\vartheta \Big[ \begin{array}{c} \beta + 2R_i^k \\ -\alpha \end{array} \Big]}{\vartheta \Big[ \begin{array}{c} 1/2 + 2R_i^k\\ -1/2 \end{array} \Big]}  \;.
\ee
\be
{\cal K}_{2}^{(k)} =  -4i \int \frac{dl}{\left(2\pi^2\right)^2}
\sum_{\alpha, \beta=0,1/2} \frac{\eta_{\alpha \beta}}{2}  \frac{\vartheta \Big[ \begin{array}{c} \beta \\ -\alpha \end{array} \Big]}{\eta^3}
 \frac{\prod_{i=1}^3 \left( - 2 \sin 2\pi R_i^k \right)^{\delta_i}}{\prod_{i=1}^3 \left( 4 \cos^2 \left(\pi R_i^k \right) \right)^{\delta_i}} \frac{\vartheta \Big[ \begin{array}{c} \beta + 2R_i^k \\ -\alpha - \theta_i^{N/2} \end{array} \Big]}{\vartheta \Big[ \begin{array}{c} 1/2 + 2R_i^k\\ -1/2 - \theta_i^{N/2} \end{array} \Big]}  \;.
\ee
Here the annulus and Klein bottle amplitudes are functions of $\tilde{q} = e^{-4\pi l}$ while the Mobius strip is a function of $-\tilde{q}$.

The RR tadpoles are given in tree channel by $\alpha=0\;, \;\beta=\half$ for the annulus and Klein bottle and $\alpha=\half\;, \;\beta=0$ for the Mobius strip.
In the UV limit $l \rightarrow \infty$ the amplitudes read
\bea
{\cal A}_{\mc{N}=1}^{(k)} &\xrightarrow[l' \rightarrow \infty]{}&- \int_{l'}^{\infty} \frac{dl}{4\pi^2} \frac14  \hbox{Tr} \left[  \gamma_{k} \right] \hbox{Tr} \left[   \gamma_{k}^{-1} \right]  \prod_{i=1}^3 \left| 2 \sin \pi \theta^k_i \right|  \;, \\
{\cal M}_{\mc{N}=1}^{(k)}  &\xrightarrow[l' \rightarrow \infty]{}& \int_{l'}^{\infty} \frac{dl}{4\pi^2}  \left[ 2  \hbox{Tr} \left[ \gamma_{\Omega'_{k}} \gamma_{\Omega'^{-T}_{k}} \right]   \prod_{i=1}^3 s_i \left( 2 \sin \left( \pi R_i^k \right) \right) \right]\;, \\
{\cal K}_{0}^{(k)} + {\cal K}_{2}^{(k)}  &\xrightarrow[l' \rightarrow \infty]{}& -\int_{l'}^{\infty} \frac{dl}{4\pi^2} 4 \left[ \prod_{i=1}^3 \left| \frac{\sin \pi R_i^k}{\cos \pi R_i^k} \right|  + \left( -1 \right)^M  \prod_{i=1}^3 \left(-1\right)^{\delta_i}\left| \frac{\sin \pi R_i^k}{\cos \pi R_i^k} \right|^{\delta_i}  \right] \;.
\eea
where $s_i = \mathrm{sgn}\left[ \sin (2\pi R_i^k) \right]$. The $(-1)^M$ factor is discussed in footnote \ref{fn:nsns}.

\section{Magnetised amplitudes}
\label{sec:magampapp}

In this appendix we calculate the Annulus and Mobius strip magnetised amplitudes. We begin with the annulus amplitudes.
The Annulus amplitudes in the background of a magnetic field is given by \cite{Conlon:2009xf}
\bea
\label{aaa}
\mc{A}_{\mc{N}=1}^{(k)} & = & -\int \frac{dt}{2t} \frac{1}{(2 \pi^2 t)}
\sum_{\alpha, \beta=0,1/2}  \frac{\eta_{\alpha \beta}}{2}
\hbox{Tr}\left( \gamma_{\theta^k} \otimes \gamma^{-1}_{\theta^k}
\frac{i(\beta_1 + \beta_2)}{2 \pi^2}
\frac{\vartheta \Big[ \begin{array}{c} \alpha \\ \beta \end{array} \Big] \left(\frac{i
\epsilon t}{2}\right)}{\vartheta \Big[ \begin{array}{c} 1/2 \\ 1/2 \end{array} \Big]
\left(\frac{i
\epsilon t}{2}\right) } \right) \nonumber  \\
& &
\times \prod_{i=1}^3
\frac{\left( - 2 \sin \pi \theta^k_i \right) \vartheta \Big[ \begin{array}{c} \alpha \\ \beta + \theta^k_i \end{array} \Big]}
{\vartheta \Big[ \begin{array}{c} 1/2 \\ 1/2 + \theta^k_i \end{array} \Big] }. \\
\label{aab1}
\mc{A}_{\mc{N}=2}^{(k)} & = & -\int \frac{dt}{2t} \frac{1}{(2 \pi^2 t)}
\sum_{\alpha, \beta=0,1/2}  \frac{\eta_{\alpha \beta}}{2} \left(-1\right)^{2\alpha}
\hbox{Tr}\left( \gamma_{\theta^k} \otimes \gamma^{-1}_{\theta^k}
\frac{i(\beta_1 + \beta_2)}{2 \pi^2}
\frac{\vartheta \Big[ \begin{array}{c} \alpha \\ \beta \end{array} \Big] \left(\frac{i
\epsilon t}{2}\right)}{\vartheta \Big[ \begin{array}{c} 1/2 \\ 1/2 \end{array} \Big]
\left(\frac{i
\epsilon t}{2}\right) } \right) \nonumber  \\
& &
\times \frac{\vartheta \Big[ \begin{array}{c} \alpha \\ \beta \end{array} \Big] }{\eta^3}\prod_{i=1}^2
\frac{\left( - 2 \sin \pi \theta^k_i \right) \vartheta \Big[ \begin{array}{c} \alpha \\ \beta + \theta^k_i \end{array} \Big]}
{\vartheta \Big[ \begin{array}{c} 1/2 \\ 1/2 + \theta^k_i \end{array} \Big] }. \\
\label{aab}
\mc{A}_{\mc{N}=4}^{(k)} & = & -\int \frac{dt}{2t} \frac{1}{(2 \pi^2 t)}
\sum_{\alpha, \beta=0,1/2}  \frac{\eta_{\alpha \beta}}{2}
\hbox{Tr}\left[ \frac{i(\beta_1 + \beta_2)}{2 \pi^2}
\frac{\vartheta \Big[ \begin{array}{c} \alpha \\ \beta \end{array} \Big] \left(\frac{i
\epsilon t}{2}\right)}{\vartheta \Big[ \begin{array}{c} 1/2 \\ 1/2 \end{array} \Big]
\left(\frac{i
\epsilon t}{2} \right) } \right] \left( \frac{\vartheta \Big[ \begin{array}{c} \alpha \\ \beta \end{array} \Big] }{\eta^3} \right)^3 \;.
\eea

The IR behaviour $t \rightarrow \infty$ of the amplitudes is used in the main part of the paper to calculate the $\beta$ functions. These can be extracted to order $B^2$ directly from the above amplitudes using the methods given in \cite{Antoniadis:1999ge,Conlon:2009xf}. In this appendix we calculate the UV behaviour. To do this we first transform to the closed string channel which gives
\bea
\mc{A}_{\mc{N}=1}^{(k)} & = & \int \frac{dl}{4\pi^2}
\sum_{\alpha, \beta=0,1/2}  \frac{\eta_{\alpha \beta}}{2}
\hbox{Tr}\left( \gamma_{\theta^k} \otimes \gamma^{-1}_{\theta^k}
\frac{i(\beta_1 + \beta_2)}{2 \pi^2}
\frac{\vartheta \Big[ \begin{array}{c} \beta \\ -\alpha \end{array} \Big] \left(\epsilon\;|\;4l\right)}{\vartheta \Big[ \begin{array}{c} 1/2 \\ -1/2 
\end{array} \Big]
\left(\epsilon\;|\;4l\right) } \right)
\nonumber  \\
& &
\times \prod_{i=1}^3
\left( - 2 \sin \pi \theta^k_i \right) \frac{ \vartheta \Big[ \begin{array}{c} \beta + \theta^k_i \\ -\alpha \end{array} \Big](4l)}
{\vartheta \Big[ \begin{array}{c} 1/2 + \theta^k_i \\ -1/2 \end{array} \Big](4l) }, \\
\mc{A}_{\mc{N}=2}^{(k)} & = & i \int \frac{dl}{4\pi^2l}
\sum_{\alpha, \beta=0,1/2}  \frac{\eta_{\alpha \beta}}{2}  \left(-1\right)^{2\alpha}
\hbox{Tr}\left( \gamma_{\theta^k} \otimes \gamma^{-1}_{\theta^k}
\frac{i(\beta_1 + \beta_2)}{2 \pi^2}
\frac{\vartheta \Big[ \begin{array}{c} \beta \\ -\alpha \end{array} \Big] \left(
\epsilon\;|\;4l\right)}{\vartheta \Big[ \begin{array}{c} 1/2 \\ -1/2 \end{array} \Big]
\left(\epsilon\;|\;4l\right) } \right)
\nonumber  \\
& &
\times \frac{\vartheta \Big[ \begin{array}{c} \beta \\ -\alpha \end{array} \Big] }{\eta^3} \prod_{i=1}^2
\left( - 2 \sin \pi \theta^k_i \right) \frac{ \vartheta \Big[ \begin{array}{c} \beta + \theta^k_i \\ -\alpha \end{array} \Big](4l)}
{\vartheta \Big[ \begin{array}{c} 1/2 + \theta^k_i \\ -1/2 \end{array} \Big](4l) }   e^{2\pi i \theta^k_i \left( \alpha-1/2 \right)}\;. \\
\mc{A}_{\mc{N}=4}^{(k)} & = & -i \int \frac{dl}{4\pi^2} \frac{1}{\left( 2l \right)^3}
\sum_{\alpha, \beta=0,1/2}  \frac{\eta_{\alpha \beta}}{2}
\hbox{Tr}\left( \frac{i(\beta_1 + \beta_2)}{2 \pi^2}
\frac{\vartheta \Big[ \begin{array}{c} \beta \\ -\alpha \end{array} \Big] \left(
\epsilon\;|\;4l\right)}{\vartheta \Big[ \begin{array}{c} 1/2 \\ 1/2 \end{array} \Big]
\left(\epsilon\;|\;4l\right) } \right)
\left( \frac{\vartheta \Big[ \begin{array}{c} \beta \\ -\alpha \end{array} \Big] }{\eta^3} \right)^3\;.
\eea
Now we can take the UV limit $l \rightarrow \infty$ which gives
\bea
\mc{A}_{\mc{N}=1}^{(k)} &\xrightarrow[l' \rightarrow \infty]{}& -\int_{l'}^{\infty}  \half \frac{dl}{4\pi^2} \hbox{Tr}\left[ \gamma_{\theta^k} \otimes \gamma^{-1}_{\theta^k}
\frac{i(\beta_1 + \beta_2)}{2 \pi^2} \left( i \left( \mathrm{cot\;}\left( \pi\epsilon\right) - \mathrm{cosec\;} \left( \pi\epsilon\right) \right) \prod_{i=1}^3  \left| 2 \sin \pi \theta^k_i \right| \right. \right.
\nonumber \\
& & \hspace{8cm} \left. \left. +  \prod_{i=1}^3  \left( - 2 \sin \pi \theta^k_i \right) \right)   \right] \;. \nonumber \\
\mc{A}_{\mc{N}=2}^{(k)}  &\xrightarrow[l' \rightarrow \infty]{}& -\int_{l'}^{\infty} \frac{dl}{4\pi^2l} \hbox{Tr}\left( \gamma_{\theta^k} \otimes \gamma^{-1}_{\theta^k}
\frac{i(\beta_1 + \beta_2)}{2 \pi^2}   i \left( \mathrm{cot\;} \left(  \pi \epsilon \right)  - \mathrm{cosec\;} \left( \pi\epsilon\right) \right) \right) \prod_{i=1}^2  \left| 2 \sin \pi \theta^k_i \right| \;. 
\nonumber  \\
\mc{A}_{\mc{N}=4}^{(k)} &\xrightarrow[l' \rightarrow \infty]{}& -\int_{l'}^{\infty}   \frac{dl}{4\pi^2} \frac{1}{\left(2l\right)^3} \hbox{Tr}\left[ \frac{i(\beta_1 + \beta_2)}{2 \pi^2} \left(  -4i \mathrm{cot\;} \left(  \pi \epsilon \right) + \frac{i\left(\mathrm{cos\;} \left(  2\pi \epsilon \right) + 3\right)}{\mathrm{sin\;} \left(  \pi \epsilon \right)}       \right) \right] \;, 
\eea
Note that these can be further simplified by using $(\beta_1 + \beta_2) \cot \left(\pi\epsilon\right) = 1 - \beta_1 \beta_2$. We can expand these expressions in powers of the magnetic field $B$ which gives up to order $B^4$ the expressions
\bea
 \mc{A}_{\mc{N}=1}^{(k)} &\xrightarrow[l' \rightarrow \infty]{}&  -\frac18  \int_{l'}^{\infty}  \frac{dl}{\left(4\pi^2\right)^2} \hbox{Tr}\left[ \gamma_{\theta^{k}} \otimes \gamma^{-1}_{\theta^{k}}
 \left( 4 \left( \beta_1 + \beta_2 \right)  \prod_{i=1}^3  \left( - 2 \sin \pi \theta^k_i \right) \right. \right. \nonumber \\
 & & \hspace{6cm} \left. \left. + \left( 4 \left(\beta_1 + \beta_2\right)^2  - \left( \beta_1^2 - \beta_2^2 \right)^2  \right) \prod_{i=1}^3  \left| 2 \sin \pi \theta^k_i \right| \right)   \right]  \;,\nonumber \\
\mc{A}_{\mc{N}=2}^{(k)} &\xrightarrow[l' \rightarrow \infty]{}&  -\frac12 \int_{l'}^{\infty} \frac{dl}{\left(4\pi^2\right)^2} \frac{1}{\left(2l\right)} \hbox{Tr}\left[ \gamma_{\theta^{k}} \otimes \gamma^{-1}_{\theta^{k}}
 \left( 4 \left(\beta_1 + \beta_2\right)^2  - \left( \beta_1^2 - \beta_2^2 \right)^2 \right)  \right]  \prod_{i=1}^2  \left| 2 \sin \pi \theta^k_i \right| \;, \nonumber \\
\mc{A}_{\mc{N}=4}^{(k)} &\xrightarrow[l' \rightarrow \infty]{}&  \int_{l'}^{\infty} \frac{dl}{\left(4\pi^2\right)^2} \frac{1}{\left(2l\right)^3}\hbox{Tr}\left[ \left( \beta_1 + \beta_2 \right)^4 \right] \;. \label{auvbexpapp}
\eea

Now we repeat the same calculations for the $\mc{N}=1$ and $\mc{N}=2$ magnetised Mobius strip amplitudes. In the open string loop channel we have
\bea
{\cal M}_{\mc{N}=1}^{(k)}  &=&     2\int_0^{\infty} \frac{dt}{2t} \frac{1}{(2 \pi^2 t)}
\sum_{\alpha, \beta=0,1/2}  \frac{\eta_{\alpha \beta}}{2}
\hbox{Tr} \left[ \frac{i}{2\pi^2} \beta \gamma_{\Omega'_{k}} \gamma_{\Omega'_{k}}^{-T} \frac{\vartheta \Big[ \begin{array}{c} \alpha \\
 \beta \end{array} \Big] \left( \frac{i\epsilon t}{2}\right)}{\vartheta \Big[ \begin{array}{c} 1/2 \\
 1/2 \end{array} \Big] \left( \frac{i\epsilon t}{2}\right) }\right]
 \prod_{i=1}^3 \left( - 2 \sin\left(\pi R_i^k \right) \right) \frac{\vartheta \Big[ \begin{array}{c} \alpha \\ \beta + R_i^k \end{array} \Big]}{\vartheta \Big[ \begin{array}{c} 1/2 \\ 1/2 +R_i^k \end{array} \Big]} \;. \nonumber \\
 {\cal M}_{\mc{N}=2}^{(k)}  &=& 2\int \frac{dt}{2t} \frac{1}{(2 \pi^2 t)}
\sum_{\alpha, \beta=0,1/2}  \frac{\eta_{\alpha \beta}}{2} \left(-1\right)^{2\alpha}
\hbox{Tr}\left( \frac{i}{2 \pi^2} \beta \gamma_{\Omega'_{k}} \gamma_{\Omega'_{k}}^{-T}
\frac{\vartheta \Big[ \begin{array}{c} \alpha \\ \beta \end{array} \Big] \left(\frac{i
\epsilon t}{2}\right)}{\vartheta \Big[ \begin{array}{c} 1/2 \\ 1/2 \end{array} \Big]
\left(\frac{i
\epsilon t}{2}\right) } \right) \nonumber  \\
& &
\times \frac{\vartheta \Big[ \begin{array}{c} \alpha \\ \beta \end{array} \Big] }{\eta^3}\prod_{i=1}^2
\frac{\left( - 2 \sin \pi R^k_i \right) \vartheta \Big[ \begin{array}{c} \alpha \\ \beta + R^k_i \end{array} \Big]}
{\vartheta \Big[ \begin{array}{c} 1/2 \\ 1/2 + R^k_i \end{array} \Big] } \;.
\eea
Again the IR behaviour can be extracted as for the Annulus. In the main section we are interested in the UV behaviour for the $\mc{N}=1$ sectors.
Transforming to the tree channel we find
\be
{\cal M}_{\mc{N}=1}^{(k)} =  -8\int \frac{dl}{4\pi^2}
\sum_{\alpha, \beta=0,1/2}  \frac{\eta_{\alpha \beta}}{2}
\hbox{Tr} \left[ \frac{i}{2\pi^2} \beta \gamma_{\Omega'_{k}} \gamma_{\Omega'^{-T}_{k}} \frac{\vartheta \Big[ \begin{array}{c} \alpha \\
 \beta \end{array} \Big] \left( \frac{\epsilon}{2}\right)}{\vartheta \Big[ \begin{array}{c} 1/2 \\
 1/2 \end{array} \Big] \left( \frac{\epsilon}{2}\right) }\right]
 \prod_{i=1}^3 \left( - 2 \sin\left(\pi R_i^k\right) \right) \frac{\vartheta \Big[ \begin{array}{c} \alpha + 2R_i^k\\ \beta +R_i^k \end{array} \Big]}{\vartheta \Big[ \begin{array}{c} 1/2 + 2R_i^k\\ 1/2 +R_i^k \end{array} \Big]} \;.
\ee
Taking the UV limit we get
\be
{\cal M}_{\mc{N}=1}^{(k)} \xrightarrow[l' \rightarrow \infty]{}  4 \int_{l'}^{\infty} \frac{dl}{4\pi^2}   \hbox{Tr} \left[ \frac{i}{2\pi^2} \beta \gamma_{\Omega'_{k}} \gamma_{\Omega'^{-T}_{k}} \left( 1 - i \left( \cot \left( \pi \epsilon/2 \right) - \frac{1}{\sin \left( \pi \epsilon/2\right)} \right)\prod_{i=1}^3 s_i \right) \right]  \prod_{i=1}^3 \left( - 2 \sin\left(\pi R_i^k\right) \right) \;.
\ee
We can expand this as
\be
{\cal M}_{\mc{N}=1}^{(k)}  \xrightarrow[l' \rightarrow \infty]{}  -4 \int_{l'}^{\infty}\frac{dl}{\left(4\pi^2\right)^2}   \hbox{Tr} \left[ \left( \beta^2 + \frac14 \beta^4\right) \gamma_{\Omega'_{k}} \gamma_{\Omega'^{-T}_{k}} \right] \prod_{i=1}^3 s_i \left( - 2 \sin\left(\pi R_i^k\right) \right) \;. 
\ee

\section{Chiral Vs Linear multiplets at 1-loop}
\label{sec:chirlin}

In this appendix we discuss the dualisation of linear multiplets to chiral multiplets in supergravity. The analysis follows that presented in \cite{Derendinger:1991hq,Binetruy:2000zx} but applied to the local IIB models studied in this paper. For more details regarding the supergravity constructions we refer to \cite{Binetruy:2000zx}.

The linear multiplet $L$ is defined by the constraint
\be
\left( D^2 - 8\bar{R} \right) L =  \left( \bar{D}^2 - 8R \right) L = 0 \;, \label{linearconst}
\ee
where $D$ is the superspace covariant derivative and $R$ is the chiral superfield containing the curvature scalar. The bosonic components are a real scalar, which we denote again by $L$, and a real two-form $\tilde{C}_2$. We can couple $L$ to a Yang-Mills (YM) gauge field ${\cal A}$ with field strength ${\cal F}$ through a Green-Schwarz coupling $\tilde{C}_2 \wedge {\cal F}$. This implies that the field strength of $\tilde{C}_2$ is modified in order to be gauge invariant under the YM gauge transformation
\be
\tilde{F} = d\tilde{C}_2 + k \Omega \;,
\ee
where $k$ is a constant and $\Omega$ is the Chern-Simons form $\Omega = \hbox{Tr} \left( {\cal A} \wedge d{\cal A}  + \frac23 {\cal A} \wedge {\cal A} \wedge {\cal A}\right)$. This modifies the Linear multiplet constraints (\ref{linearconst}) to
\be
\left( D^2 - 8\bar{R} - 2k \hbox{Tr} {\cal W}^2  \right) L =  \left( \bar{D}^2 - 8R  - 2k \hbox{Tr} {\cal \bar{W}}^2 \right) L = 0 \;, \label{gaugelinearconst}
\ee
with $\hbox{Tr}{\cal W}^2 = \frac12 \left(\bar{D}^2 - 8R\right)\Omega$ the usual YM field strength.  The supergravity action for single linear multiplet $L$ and chiral multiplets (collectively denoted as) $T$ is given by
\bea
S = -3 \int E F\left(T,\bar{T},L \right) + \left\{ \frac12 \int \frac{E}{R} e^{K/2} W(T) + \mathrm{hc}\right\} + ... \;,
\eea
where $E$ is the super-vielbein, $W$ the superpotential, and $K\left(T,\bar{T}, L\right)$ is the Kahler potential. The function $F\left(T,\bar{T},L\right)$ is called the {\it subsidiary function} and is related to the Kahler potential through
\be
F - L F_L = 1 - \frac13 L K_L \;, \label{normconst}
\ee
where subscripts denote derivatives.
Equation (\ref{normconst}) comes from the fact that in the linear multiplet formalism integrating over the superdeterminant gives a non-canonically normalised Einstein term and (\ref{normconst}) is the correct normalisation constraint. The constraint (\ref{normconst}) fixes
\be
F\left( T,\bar{T},L\right) = 1 + L \Delta\left(T,\bar{T}\right) + \frac{L}{3} \int \frac{d\lambda}{\lambda} K_{\lambda} \left( T,\bar{T},\lambda\right) \;, \label{genF}
\ee
with $\lambda$ a dummy variable for $L$. $\Delta\left(T,\bar{T}\right)$ is call the {\it Linear potential} and forms part of the gauge coupling of $L$ to ${\cal A}$.
which is given by \cite{Binetruy:2000zx}
\be
3k\left(\frac{F\left(T,\bar{T},L \right)-1}{L} \right) \hbox{Tr} {\cal F}^2 \;. \label{lineargauge}
\ee
In our case $\Delta\left(T,\bar{T}\right)$ will encode the 1-loop correction to the gauge coupling associated to the field $L$.

We wish to dualise the linear multiplet $L$ to a chiral multiplet $M$ (which has a propagating bosonic component of single complex scalar field $M$). The $\hbox{Re}M$ is dual to the scalar $L$ and $\hbox{Im}M$ is dual to $\tilde{C}_2$. To do this we introduce the coupling to the action
\be
S = -3 \int E \left[ F\left(T,\bar{T},L \right) + \left( L - k\Omega \right) \left( M + \bar{M} \right)  \right]  + ... \;. \label{chiralact}
\ee
Then we find the equation of motion for $L$ which read\footnote{Note that $\delta_L E = -\frac13 E K_L \delta L$ and $\delta_L \Omega = \frac13 \Omega K_L \delta L$. }
\be
\left(M+\bar{M}\right) \left( 1 - \frac13 L K_L \right) = \frac13 F K_L - F_L \;. \label{eqmol}
\ee
This equation can be used in principle to solve for $L\left( M+\bar{M}, T, \bar{T} \right)$ and write the action using only chiral superfields. Now using (\ref{normconst}) and (\ref{eqmol}) we find
\be
F\left(T,\bar{T},L \right) + L\left(M+\bar{M} \right) = 1 \;. \label{MfromF}
\ee
Substituting this into the action (\ref{chiralact}) gives
\bea
S &=& -3\int E \left[ 1 - k\Omega\left(M+\bar{M} \right)\right] + ... \nonumber \\
    &=& -3\int E - \left\{ \frac38 k \int \frac{E}{R} M \left( {\cal \bar{D}}^2 - 8R \right) \Omega + \mathrm{hc} \right\} + ... \nonumber \\
    &=& -3\int E - \left\{ \frac34 k \int \frac{E}{R} M \hbox{Tr}{\cal W}^2 + \mathrm{hc} \right\} + ... \;.
\eea
Here in the first step the derivative terms vanish upon integration by parts and we used the expression below (\ref{gaugelinearconst}) in the second step.
This implies that in the chiral multiplet formalism the familiar holomorphic gauge kinetic function $f(M)$ takes the form
\be
f = -6kM \;. \label{chiralgauge}
\ee
Equivalently this can be derived by simply substituting (\ref{MfromF}) into (\ref{lineargauge}) and using holomorphy.
This should be compared with what is denoted the tree-level gauge kinetic function in the main text (\ref{treegauge}). We see that they match and so what remains is to determine the precise relationship between $\hbox{Re}M$ and $L$.

\subsection{Application to local IIB models}

In this subsection we apply the results derived in the previous section to the local IIB models studied in the main text. Of course the analysis of the previous section is vastly over simplified since there are many fields in a concrete construction but the key properties can be deduced considering only one field which is what we do in this subsection.

We begin by specifying the Kahler potential which in local models takes the form
\be
K = -2 \mathrm{ln} {\cal V}\left(T,\bar{T}\right) + {\cal V}\left(T,\bar{T}\right) L^2 + K_{0}\left(U,S\right) \;, \label{localkahlerpot}
\ee
where $U$ and $S$ are the complex-structure and dilaton fields respectively and ${\cal V}\left(T,\bar{T}\right)$ denotes the CY volume as a function of the four-cycle volumes $T$.\footnote{The factor of ${\cal V}$ in front of the $L^2$ term is justified from the fact that after dualisation to the chiral multiplet this Kahler potential term takes the form $K\left(M,\bar{M}\right) = \frac{9\left(M+\bar{M}\right)^2}{4{\cal V}}$ which is the form advocated in \cite{08105660} for the blow-up modulus.} This Kahler potential, using (\ref{genF}), gives a subsidiary function of
\be
F = 1 + L\Delta + \frac{2{\cal V}}{3}L^2 \;. \label{localsub}
\ee
Then using (\ref{localkahlerpot}) and (\ref{MfromF}) gives
\bea
\hbox{Re}M &=& -\half \left( \frac{2{\cal V}}{3} L + \Delta \right) \;.
\eea
This should be compared with (\ref{redef2}) and we see that they match up to a field rescaling $m=\frac{-{\cal V}}{3}L$. Using (\ref{chiralgauge}) we can also match the field coupling $-6k=s$.

Finally we need to determine the nature of $\Delta$. From (\ref{localsub}) we see that at the orbifold limit $L=0$ the gauge coupling (\ref{lineargauge}) takes the form
\be
\frac{1}{g^2} = 3k\Delta \;.
\ee
This shows that the field redefinition is given by the 1-loop correction to the gauge coupling as required in the main text.

\section{Some conventions and formulae}
\label{confor}

We here collate definitions and identities of the various Jacobi-$\vartheta$ functions.
We write $q = e^{-\pi t}$ throughout these formulae. The eta function is defined by
\be
\eta(t) = q^{1/24} \prod_{n=1}^{\infty}(1 - q^n).
\ee
The Jacobi $\vartheta$-function with general characterstic is defined as
\be
\vartheta \Big[ \begin{array}{c} \alpha \\ \beta \end{array} \Big](z | t) =
\sum_{n \in \, \mbb{Z}} q^{(n + \alpha)^2/2} e^{2 \pi i (z + \beta) (n + \alpha)}.
\ee
Here $z = 0$ unless specified.
The $\vartheta$ functions are manifestly invariant under $\alpha \to \alpha + \mbb{Z}$.
A useful expansion valid for $\alpha \in (-\half, \half]$ is
\be
\frac{\vartheta \Big[ \begin{array}{c} \alpha \\ \beta \end{array} \Big]}{\eta}(t)
= e^{2 \pi i \alpha \beta} q^{\frac{\alpha^2}{2} - \frac{1}{24}} \prod_{n=1}^{\infty}
\left( 1 + e^{2 \pi i \beta} q^{n- \half + \alpha} \right) \left(1 + e^{-2 \pi i \beta} q^{n-\half -\alpha}  \right).
\ee
For the four special $\vartheta$-functions, we have
\bea
\vartheta_1(z | t) \equiv \vartheta \Big[ \begin{array}{c} \half \\ \half \end{array} \Big](z | t)
& = & 2 q^{1/8} \sin \pi z \prod_{n=1}^{\infty} (1 - q^n) (1-e^{2 \pi iz} q^n) (1 - e^{-2 \pi i z} q^n). \\
\vartheta_2(z | t) \equiv \vartheta \Big[ \begin{array}{c} \half \\ 0 \end{array} \Big](z | t)
& = & 2 q^{1/8} \cos \pi z \prod_{n=1}^{\infty} (1 - q^n) (1 + e^{2 \pi iz} q^n) (1 + e^{-2 \pi i z} q^n). \\
\vartheta_3(z | t) \equiv \vartheta \Big[ \begin{array}{c} 0 \\ 0 \end{array} \Big](z | t)
& = &  \prod_{n=1}^{\infty} (1 - q^n) (1 + e^{2 \pi iz} q^{n-\half}) (1 + e^{-2 \pi i z} q^{n-\half}). \\
\vartheta_4(z | t) \equiv \vartheta \Big[ \begin{array}{c} 0 \\ \half \end{array} \Big](z | t)
& = &  \prod_{n=1}^{\infty} (1 - q^n) (1 - e^{2 \pi iz} q^{n-\half}) (1 - e^{-2 \pi i z} q^{n-\half}).
\eea

The functions transform as (from \cite{Antoniadis:1999ge})
\bea
 \vartheta \Big[ \begin{array}{c} \alpha \\ \beta \end{array} \Big](z | \tau) &=& e^{-2\pi i \alpha \beta - \frac{i\pi z^2}{\tau}} \sqrt{\frac{i}{\tau}} \vartheta \Big[ \begin{array}{c} -\beta \\ \alpha \end{array} \Big] \left( \frac{z}{\tau} | -\frac{1}{\tau} \right)   \;, \\
 \vartheta \Big[ \begin{array}{c} \alpha \\ \beta \end{array} \Big](z | \tau) &=& e^{\pi i \alpha \left( \alpha - 1\right)}  \vartheta \Big[ \begin{array}{c} \alpha \\ \beta -\alpha +  1/2 \end{array} \Big](z | \tau + 1)
\;. \label{modtrathe}
\eea

\end{document}